\RequirePackage{lineno}
\documentclass[%
 reprint,
 amsmath,amssymb,
 aps,twocolumn,
]{revtex4}

\usepackage{graphicx}
\usepackage{dcolumn}
\usepackage{bm}
\usepackage{amsmath}
\usepackage{lineno}
\usepackage{xspace}
\usepackage{siunitx}
\usepackage{ulem}
\usepackage{color}
\usepackage{xcolor}
\usepackage{longtable}
\usepackage{orcidlink}

\DeclareRobustCommand{\erase}{\bgroup\markoverwith{\textcolor{red}{\rule[.5ex]{2pt}{0.4pt}}}\ULon}
\DeclareRobustCommand{\eraseblue}{\bgroup\markoverwith{\textcolor{blue}{\rule[.5ex]{2pt}{0.4pt}}}\ULon}
\def\b0b {$\overset{\textbf{\fontsize{2pt}{2pt}\selectfont(---)}}{B^{0}}$}

\begin{document}

\preprint{APS/123-QED}

\title{Measurement of branching fractions, {\it CP} asymmetry, and isospin asymmetry for $\boldsymbol{B\rightarrow\rho\gamma}$ decays using Belle and Belle~II data}
  \author{I.~Adachi\,\orcidlink{0000-0003-2287-0173}} 
  \author{K.~Adamczyk\,\orcidlink{0000-0001-6208-0876}} 
  \author{L.~Aggarwal\,\orcidlink{0000-0002-0909-7537}} 
  \author{H.~Aihara\,\orcidlink{0000-0002-1907-5964}} 
  \author{N.~Akopov\,\orcidlink{0000-0002-4425-2096}} 
  \author{A.~Aloisio\,\orcidlink{0000-0002-3883-6693}} 
  \author{N.~Anh~Ky\,\orcidlink{0000-0003-0471-197X}} 
  \author{D.~M.~Asner\,\orcidlink{0000-0002-1586-5790}} 
  \author{H.~Atmacan\,\orcidlink{0000-0003-2435-501X}} 
  \author{T.~Aushev\,\orcidlink{0000-0002-6347-7055}} 
% ...
\collaboration{The Belle and Belle II Collaborations}

\date{\today}

\begin{abstract}
We present measurements of $B^{+}\rightarrow\rho^{+}\gamma$ and $B^{0}\rightarrow\rho^{0}\gamma$ decays
using a combined data sample of $772 \times 10^6$ $B\overline{B}$ pairs collected by the Belle experiment
and $387\times 10^6$ $B\overline{B}$ pairs collected by the Belle~II experiment
in $e^{+}e^{-}$ collisions at the $\Upsilon (4S)$ resonance.
After an optimized selection, a simultaneous fit to the Belle and Belle~II data sets yields
$114\pm 12$ $B^{+}\rightarrow\rho^{+}\gamma$ and $99\pm 12$ $B^{0}\rightarrow\rho^{0}\gamma$ decays.
The measured branching fractions are $(13.1^{+2.0 +1.3}_{-1.9 -1.2})\times 10^{-7}$ and $(7.5\pm 1.3^{+1.0}_{-0.8})\times 10^{-7}$ for $B^{+}\rightarrow\rho^{+}\gamma$ and $B^{0}\rightarrow\rho^{0}\gamma$ decays, respectively,
where the first uncertainty is statistical and the second is systematic.
We also measure the isospin asymmetry $A_{\rm I}(B\rightarrow\rho\gamma)=(10.9^{+11.2 +7.8}_{-11.7 -7.3})\%$ and
the direct {\it CP} asymmetry $A_{\it CP}(B^{+}\rightarrow\rho^{+}\gamma)=(-8.2\pm 15.2^{+1.6}_{-1.2})\%$.
\end{abstract}

\maketitle


\section{Introduction}
Flavor-changing neutral-current (FCNC) processes are sensitive probes of physics beyond the standard model (SM).
These decays are forbidden at tree level, making them particularly sensitive to beyond-the-SM contributions, which could be significant compared to loop-level SM amplitudes.
Beyond-the-SM physics searches with radiative decays of $B$ mesons, such as $B\rightarrow K^{\ast}\gamma$ and $B\rightarrow\rho\gamma$, is attractive,
as these decays are FCNC processes dominated by the $b\rightarrow (s,d)\gamma$ transition, which makes it easier to distinguish beyond-the-SM contributions~\cite{Belle-II:2018jsg}.

The $B\rightarrow\rho\gamma$ decay involves a $b\rightarrow d$ transition at the quark level and, within the SM, has a branching fraction one order of magnitude smaller than radiative $B$ decays involving $b\rightarrow s$ transitions.
However, the $B\rightarrow\rho\gamma$ decay mode can be affected by beyond-the-SM processes not appearing in $b\rightarrow s$ processes.
The decay $B\rightarrow\rho\gamma$ was first observed by Belle experiment~\cite{Belle:2005grh} in 2006 and later confirmed by BaBar experiment~\cite{BaBar:2006lms}.
Both experiments subsequently made more precise measurements with larger data sets:
$657\times 10^{6}~B\overline{B}$ pairs for Belle~\cite{Belle:2008imr} and $465\times 10^{6}~B\overline{B}$ pairs for BaBar~\cite{BaBar:2008txc}.
In the SM, the $B\rightarrow\rho\gamma$ {\it CP}-averaged isospin asymmetry ($\bar{A}_{\rm I}^{\rm SM}$) is predicted to be close to that of $B\rightarrow K^{\ast}\gamma$ and to equal $(5.2\pm 2.8)\%$~\cite{Lyon:2013gba}.
This asymmetry is defined as
$\bar{A}_{\rm I}=\left(A_{\rm I}^{b}+A_{\rm I}^{\bar{b}}\right)/2$, with
\begin{linenomath}
 \begin{align}
  A_{\rm I}^{b}=\frac{c^{2}_{\rho}\Gamma\left(\overline{B}{}^{0}\rightarrow\rho^{0}\gamma\right)-\Gamma\left(B^{-}\rightarrow\rho^{-}\gamma\right)}{c^{2}_{\rho}\Gamma\left(\overline{B}{}^{0}\rightarrow\rho^{0}\gamma\right)+\Gamma\left(B^{-}\rightarrow\rho^{-}\gamma\right)},
 \end{align}
\end{linenomath}
and $A_{\rm I}^{\bar{b}}$ defined similarly but with the {\it CP}-conjugate modes.
The factor $c_{\rho}= \sqrt{2}$ results from the quark content of the $\rho^{0}$ meson (i.e., a Clebsch-Gordan coefficient).
Experimentally, only the isospin asymmetry with {\it CP}-averaged branching fractions, $A_{\rm I}$, has been measured to date:
\begin{linenomath}
 \begin{align}
  A_{\rm I}=\frac{ c_{\rho}^{2}\Gamma(\overset{\textbf{\fontsize{2pt}{2pt}\selectfont(---)}}{B^{0}}\rightarrow\rho^{0}\gamma)-\Gamma(B^{\pm}\rightarrow \rho^{\pm}\gamma) }{ c_{\rho}^{2}\Gamma(\overset{\textbf{\fontsize{2pt}{2pt}\selectfont(---)}}{B^{0}}\rightarrow\rho^{0}\gamma)+\Gamma(B^{\pm}\rightarrow \rho^{\pm}\gamma) },
 \end{align}
\end{linenomath}
which equals $\bar{A}_{I}$ if the {\it CP} asymmetry
\begin{linenomath}
 \begin{align}
  A_{\it CP}(B\rightarrow\rho\gamma)=\frac{\Gamma\left(\overline{B}\rightarrow\overline{\rho}\gamma\right)-\Gamma\left(B\rightarrow\rho\gamma\right)}{\Gamma\left(\overline{B}\rightarrow\overline{\rho}\gamma\right)+\Gamma\left(B\rightarrow\rho\gamma\right)}
 \end{align}
\end{linenomath}
is the same for $B^{+}\rightarrow\rho^{+}\gamma$ and $B^{0}\rightarrow\rho^{0}\gamma$ decays.
The current world average of isospin asymmetry measurements $A_{\rm I}^{\mathrm{exp}}(\rho\gamma)=(30^{+16}_{-13})\%$~\cite{ParticleDataGroup:2022pth} is about two standard deviations from the SM expectation.

Here we report measurements of $B\rightarrow\rho\gamma$ decays performed using both Belle and Belle~II data sets.
The signal decay modes are $B^{+}\rightarrow\rho^{+}\left(\rightarrow\pi^{0}\pi^{+}\right)\gamma$
and $B^{0}\rightarrow\rho^{0}\left(\rightarrow\pi^{+}\pi^{-}\right)\gamma$~\cite{charge-conjugate}.
We use the full Belle data set corresponding to 711~${\mathrm {fb}^{-1}}$ taken at the $\Upsilon(4S)$ resonance energy (on-resonance) containing $(772\pm 11)\times 10^6$ $B\overline{B}$ pairs.
In addition, we use a 362~${\mathrm {fb}^{-1}}$ Belle~II data set collected from 2019--2022, containing $(387\pm 6)\times 10^6$ $B\overline{B}$ pairs.
We also use off-resonance data sets,
collected at an energy 60~MeV below the $\Upsilon (4S)$ resonance,
to study continuum background ($e^+e^-\rightarrow q\overline{q}$, where $q=u,d,s,c$).
The off-resonance data samples correspond to 89.5~${\rm fb^{-1}}$ and 42.3~${\rm fb^{-1}}$ for Belle and Belle~II, respectively.
After applying selection criteria to identify $B\to\rho\gamma$ candidates, we fit the data to determine 
$B^0$ and $B^+$ branching fractions and asymmetries $A^{}_I$ 
and $A^{}_{CP}$.


This paper is organized as follows.
Section~\ref{sec:detector} introduces the Belle and Belle~II detectors followed by the description of the data and simulated samples.
The event selection and reconstruction of the decays are described in section~\ref{sec:selection}.
The methods to suppress background are presented in section~\ref{sec:cut}.
Section~\ref{sec:fit} explains the fitting procedure to extract signal yields.
The systematic uncertainties are discussed in section~\ref{sec:systematics}.
The results are presented in section~\ref{sec:results}. 

\section{Detectors and data sets}
\label{sec:detector}
The Belle detector~\cite{Belle:2000cnh,PTEP_belle} was a large-solid-angle spectrometer that operated at the KEKB asymmetric-energy $e^{+}e^{-}$ (3.5~GeV on 8.0~GeV) collider~\cite{Kurokawa:2001nw,PTEP_kekb}.
The detector consisted of a silicon vertex detector and a central drift chamber (CDC) for reconstructing trajectories of charged particles (tracks), an array of aerogel Cherenkov counters (ACC) and time-of-flight scintillation counters (TOF) for identifying charged particles,
and an electromagnetic CsI(Tl) crystal calorimeter (ECL) for identifying photons and electrons.
These detectors were surrounded by a superconducting solenoid coil providing a magnetic field of 1.5~T.
An iron flux return yoke located outside the coil was instrumented with resistive-plate chambers to detect $K^{0}_{L}$ mesons and identify muons.

The Belle~II detector~\cite{Belle-II:2010dht} is an upgrade of the Belle detector and operates at the SuperKEKB $e^{+}e^{-}$ collider.
The energies of electron and positron beams are 7.0~GeV and 4.0~GeV, respectively.
The vertex detector consists of pixel sensors and double-sided silicon strips.
The Belle~II CDC is surrounded by two types of Cherenkov light detector systems used for particle identification:
a time-of-propagation detector (TOP) for the barrel region (32.2$^{\circ}$ to 128.7$^{\circ}$),
and an aerogel ring-imaging Cherenkov detector (ARICH) for the forward endcap region (12.4$^{\circ}$ to 31.4$^{\circ}$).
The Belle ECL is re-used in Belle~II along with the solenoid and the iron flux return yoke.
For both Belle and Belle~II, the $z$ axis is defined as the central axis of the solenoid,
with the positive direction being very close to the direction of the electron beam.

Monte Carlo simulated samples are used to optimize selection criteria, study sources of backgrounds, calculate reconstruction efficiencies, and determine probability density functions (PDFs) used for fitting the data.
We use {\sc EvtGen}~\cite{Lange:2001uf} to generate $e^{+}e^{-}\rightarrow B\overline{B}$ pairs.
In addition, we use {\sc Pythia} 8.2~\cite{Sjostrand:2014zea} for the Belle~II sample.
We simulate continuum background events using {\sc Pythia} 6.4~\cite{Sjostrand:2006za} for the analysis of Belle data, and KKMC~\cite{Ping:2008zz} for hard scattering followed by {\sc Pythia} 8.2~\cite{Sjostrand:2014zea} for hadronization in Belle~II.
We simulate the effects of final state radiation by {\sc Photos}~\cite{photos} for both Belle and Belle~II.
The detector response is based on {\sc geant3}~\cite{geant3} and {\sc geant4}~\cite{geant4} for Belle and Belle~II data, respectively.
We use simulated samples of ${\rm 2\times 10^{6}}$ signal events and more than ${\rm 1ab^{-1}}$ of background events for our studies.
The data and simulation are processed using the Belle~II analysis software framework~\cite{basf2,basf2_repo,Gelb:2018agf}.

\section{Event selection}
\label{sec:selection}
Most of the selection criteria used for Belle and Belle~II are similar; however, certain criteria are adjusted to account for the improved performance of the upgraded detector.

The triggers of Belle and Belle~II use either the number of tracks, or ECL energy depositions (clusters), or the total energy of all clusters.
The trigger efficiency for signal events is almost 100\%.
In the offline analysis, a high energy primary photon ($\gamma_{\rm prim}$) candidate is reconstructed from an ECL cluster not associated with any track.
Only ECL clusters whose polar angle $\theta$ is in the barrel region are considered.
The c.m.\ energy of the primary photon candidate ($E_{\gamma}^{\ast}$) is required to satisfy ${\rm 1.8~GeV}<E_{\gamma}^{\ast}<{\rm 2.8~GeV}$.
For Belle, we require $E_{\rm 9}/E_{\rm 25}>0.95$ to select a cluster shape consistent with an electromagnetic shower,
where $E_{\rm 9}/E_{\rm 25}$ is defined as the ratio of energy deposits within a $3\times 3$ array of CsI(Tl) crystals and within a $5\times 5$ array, both centered on the crystal with the highest energy.
For Belle~II, we require $E_{\rm 9}/E_{\rm 21}>0.95$,
where $E_{21}$ is defined similarly to $E_{25}$ but with energy deposits at the four corners of the $5\times 5$ array removed.
We also define for Belle~II a cluster second moment $S\equiv \Sigma_{i} w_{i}E_{i}r_{i}^{2} / \Sigma_{i} w_{i}E_{i}$,
where $E_{i}$ is the energy deposited in the $i$th crystal,
and $r_{i}$ is the distance in the plane perpendicular to the shower axis
from the center of the cluster to the center of the $i$th crystal.
The energies are weighted by factors $w_{i}$ ranging from 0.0--1.0 that account for a crystal energy being shared among overlapping clusters.
The sum is taken over all crystals in a cluster.
To reject clusters resulting from hadron showers,
we require that there be at least eight crystals in a cluster.
To better identify the primary photon cluster,
we require $S<{\rm 1.5}~{\rm cm^{2}}$.

Track candidates are required to satisfy $dr<0.5$~cm and $|dz|<2$~cm,
where $dr$ and $dz$ are transverse and longitudinal impact parameters, respectively.
We select $\pi^{+}$ candidates from tracks by requiring ${\cal R}^{}_{\pi/K} = {\cal L}^{}_{\pi} / ({\cal L}^{}_{\pi} + {\cal L}^{}_K )>0.6$,
where ${\cal L}^{}_{\pi (K)}$ is the likelihood for
a pion (kaon) hypothesis.
These likelihoods are based on information from the CDC, ACC, and TOF detectors in Belle,
and from the CDC, TOP, and ARICH detectors in Belle~II.

Neutral $\pi^{0}$ mesons are reconstructed via $\pi^{0}\rightarrow\gamma\gamma$ decays.
In Belle data, $\pi^{0}$ mesons are reconstructed using two photons, each with energy greater than 50~MeV.
We require the invariant mass of the two photons to satisfy ${\rm 119~MeV}/c^{2}<M(\gamma\gamma)<{\rm 151~MeV}/c^{2}$.
In Belle~II data, the energy thresholds of the daughter photons depend on the region in the ECL.
For the barrel region, the daughter photons are required to have energies greater than 30~MeV,
while for the forward and backward endcap regions we use thresholds of 80~MeV and 60~MeV, respectively.
In addition, the daughter photons must satisfy the condition $\Sigma w_{i}>{\rm 1.5}$.
The invariant mass of the two photons in Belle~II is required to be in the range ${\rm 120~MeV}/c^{2}<M(\gamma\gamma)<{\rm 145~MeV}/c^{2}$.

We reconstruct $\rho^{0}\rightarrow\pi^{+}\pi^{-}$ and $\rho^{+}\rightarrow\pi^{+}\pi^{0}$ candidates
by requiring that the invariant mass of the pion pair be in the range ${\rm [0.64, 0.90]~GeV}/c^{2}$ for Belle data,
and the range ${\rm [0.65, 0.90]~GeV}/c^{2}$ for Belle~II data.
A $B$ candidate is subsequently reconstructed by combining a $\rho$ candidate with a $\gamma_{\rm prim}$ candidate.

To remove low multiplicity and $q\overline{q}$ events, we require at least two tracks in the event and that the ratio of the second to zeroth Fox-Wolfram moments~\cite{fox} be less than 0.7.
To fit for the signal yield, we define two observables:
the beam-energy-constrained mass $M_{\rm bc}\equiv\sqrt{\left(E_{\rm beam}^{\ast}/c^{2}\right)^2 - |{\bf p}_{B}^{\ast}/c|^2}$,
and the energy difference $\Delta E\equiv E_{B}^{\ast}-E_{\rm beam}^{\ast}$,
where $E_{\rm beam}$ is the beam energy and $E_{B}$ and $p_{B}$ are the reconstructed energy and momentum, respectively, of the signal $B$ candidate.
All quantities are calculated in the $e^{+}e^{-}$ c.m. frame.
For the $B^{0}\rightarrow\rho^{0}\gamma$ mode, we improve the resolution in $M_{\rm bc}$ by substituting, for the magnitude of the photon momentum,
the difference between the beam energy and the energy of the $\rho^{0}$ candidate:
$\mbox{\boldmath $\vec{p}$}_{B^0}\rightarrow \mbox{\boldmath $\vec{p}$}_{\rho^{0}} + \left(\mbox{\boldmath $\vec{p}$}_{\gamma} / |\mbox{\boldmath $\vec{p}$}_{\gamma}|\right) \left(E_{\rm beam}-E_{\rho^{0}}\right)/c$,
where $E_{\rho^{0}}$ and $\mbox{\boldmath $\vec{p}$}_{\rho^{0}}$ are the energy and momentum, respectively, of the $\rho^{0}$ candidate, 
and $\mbox{\boldmath $\vec{p}$}_{\gamma}$ is the momentum of 
the $\gamma_{\rm prim}$ candidate.
According to MC simulation, this modification reduces the width of the $M_{\rm bc}$ distribution by 18\%.
We retain signal candidates that satisfy $M_{\rm bc}>5.2~{\rm GeV/}c^{2}$ and $|\Delta E|<0.3~{\rm GeV}$.
In addition, we define a variable $M(K\pi)$, which is the invariant mass of the $\rho$ candidate calculated assuming that one of the charged pions is a kaon.
For $\rho^{0}\rightarrow\pi^{+}\pi^{-}$ decays, the substitution of the kaon mass for the pion mass is applied to the pion with the lower value of ${\cal R^{}_{\pi/{\it K}}}$.
This allows us to distinguish $B\rightarrow\rho\gamma$ from $B\rightarrow K^{\ast}\gamma$, which peaks in the $K^{\ast}\left(892\right)^0$ mass region.
We retain events satisfying $M(K\pi)\in \left(0.80, 1.50\right)~{\rm GeV}/c^{2}$.
To reduce combinatorial background due to $B\overline{B}$ events, the $B$ meson candidate must also satisfy $|\cos{\theta_{h}}|<0.8$,
where $\theta_{h}$ is the angle in the $B$ rest frame between the momentum of $\gamma_{\rm prim}$
and the negative of the boost direction of the laboratory frame.
For correctly reconstructed signal decays, the $|\cos{\theta_{h}}|$ distribution is uniform,
while for combinatorial background, the distribution tends to peak at $|\cos{\theta_{h}}|\approx 1$.
After applying the above selection criteria, 12.3\% (3.0\%) of events have multiple $B^{+}$ ($B^{0}$) candidates.
For such multiple candidate events, the average multiplicity is 2.3 for both $B^{+}$ and $B^{0}$ decays.
To minimize potential bias, we select one candidate randomly.

To assess the quality of the simulation, we compare its predictions with data using two control samples,
$B\rightarrow \overline{D}\pi^{+}$ and $B\rightarrow K^{\ast}\gamma$ decays.
We reconstruct both $\overline{D}{}^{0}\rightarrow K^{+}\pi^{-}$ and $D^{-}\rightarrow K^{+}\pi^{-}\pi^{-}$ decays, requiring that the invariant masses be in the ranges ${\rm 1.85~GeV}/c^{2}<M(K^{+}\pi^{-})<{\rm 1.88~GeV}/c^{2}$ and ${\rm 1.86~GeV}/c^{2}<M(K^{+}\pi^{-}\pi^{-})<{\rm 1.88~GeV}/c^{2}$.
The $K^{\ast}$ mesons are reconstructed via $K^{\ast +}\rightarrow K^{+}\pi^{0}$ and $K^{\ast 0}\rightarrow K^{+}\pi^{-}$ decays, where the charged kaon is required to have ${\cal R}^{}_{K/\pi}>0.6$
and the $\pi^{0}$ and $\pi^{-}$ must satisfy the same criteria as applied to signal $B\rightarrow\rho^{0}\gamma$ decays.
The invariant mass of the $K\pi$ meson pair must satisfy ${\rm 0.817~GeV}/c^{2}<M(K^{+}\pi^{-(0)})<{\rm 0.967~GeV}/c^2$.

\section{Background suppression}
\label{sec:cut}
After applying the above selection criteria, the main remaining source of background is $q\overline{q}$ events,
where the candidate $\gamma_{\rm prim}$ results from an asymmetric $\pi^{0}\rightarrow\gamma\gamma$ or $\eta\rightarrow\gamma\gamma$ decay.
In this case, one of the photons has much higher energy than the other.
To reduce this background, we use two boosted decision trees:
a dedicated $\pi^{0}(\eta)$ veto and one for $q\overline{q}$ suppression.

\subsection{$\pi^{0}(\eta)$ veto}
\label{sec:pi0etaveto}
For the $\pi^{0}$ and $\eta$ veto, we pair $\gamma_{\rm prim}$ with other photons in the event to reconstruct $\pi^{0}\rightarrow\gamma\gamma$ and $\eta\rightarrow\gamma\gamma$ decays.
If such a candidate is identified, the $\gamma_{\rm prim}$ candidate is removed.
The energy of the photon with which $\gamma_{\rm prim}$ is paired ($\gamma_{\rm soft}$) must exceed a polar-angle-dependent threshold.
For Belle~II data,
we require that the sum of crystal weights $w_{i}$ for the $\gamma_{\rm soft}$ cluster be at least two, and that this cluster be reconstructed in time with the event trigger.
The probability of a correctly reconstructed $\pi^{0}(\eta)$ is calculated using a boosted decision tree ({$BDT_{v}$}) trained using simulated events.
The $BDT_{v}$ uses up to eight input variables:
 (1) the invariant mass of the two photons $M(\gamma_{\rm prim}\gamma_{\rm soft})$;
 (2) the $\gamma_{\rm soft}$ energy;
 (3) the $\gamma_{\rm soft}$ polar angle;
 (4) $\Sigma w_{i}$ for the $\gamma_{\rm soft}$ cluster;
 (5) the $\gamma_{\rm soft}$ $E_{9}/E_{25(21)}$;
 (6) the cosine of the angle in the $\pi^{0}(\eta)$ rest frame between the momentum of $\gamma_{\rm prim}$ and the negative of the boost direction of the laboratory frame;
 (7) the distance between the ECL cluster of $\gamma_{\rm soft}$ and the nearest track extrapolated to the ECL; and
 (8) the output of a multivariate classifier based on Zernike moments of the $\gamma_{\rm soft}$ shower~\cite{zernike}. These moments are defined in the plane perpendicular to the shower direction.
Requirements (7) and (8) are not applied to Belle data, while for Belle~II data, variables (4), (5), and (8) are used only for the $\eta$ veto.
The {$BDT_{v}$} value is calculated for all $\gamma_{\rm soft}$ candidates. Among all $\gamma_{\rm prim}\gamma_{\rm soft}$ pairs, the one with the highest {$BDT_{v}$} value is selected.
The background in which the $\gamma_{\rm prim}$ originates from $\pi^{0}(\eta)\rightarrow\gamma\gamma$ decays tends to peak at one.
Selection criteria on the {$BDT_{v}$} output are optimized to maximize the signal significance, $n_{s}/\sqrt{n_{s}+n_{b}}$, where $n_{s(b)}$ is the number of simulated signal (background) events after the {$BDT_{v}$} selection
in a signal-enriched region defined as $M_{\rm bc}>{\rm 5.27~GeV}/c^2$, ${\rm -0.20~GeV}<\Delta E<{\rm 0.10~GeV}$, and $M(K\pi)>{\rm 0.92~GeV}/c^2$.
The optimized selection for $BDT_{v}$ rejects about 75\% (82\%) of $q\overline{q}$ background for Belle (Belle~II) data, while retaining about 83\% (89\%) of signal decays.

\begin{figure}[h]
 \includegraphics[width=90mm]{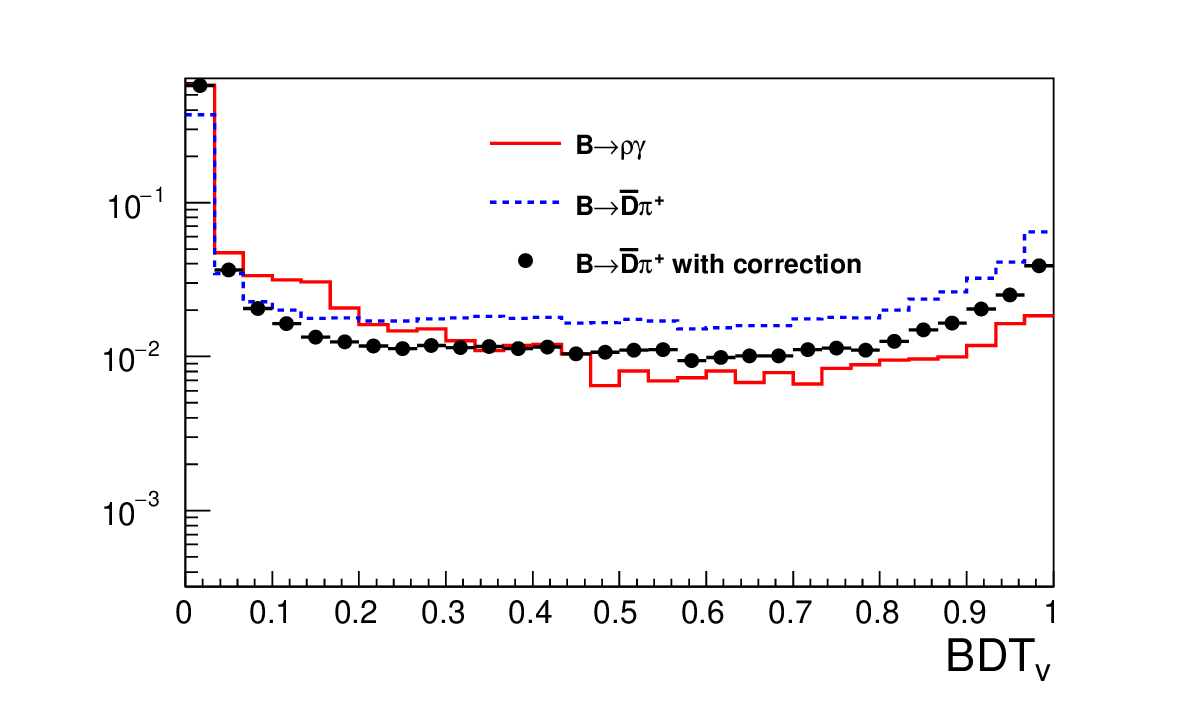}\\
 \includegraphics[width=90mm]{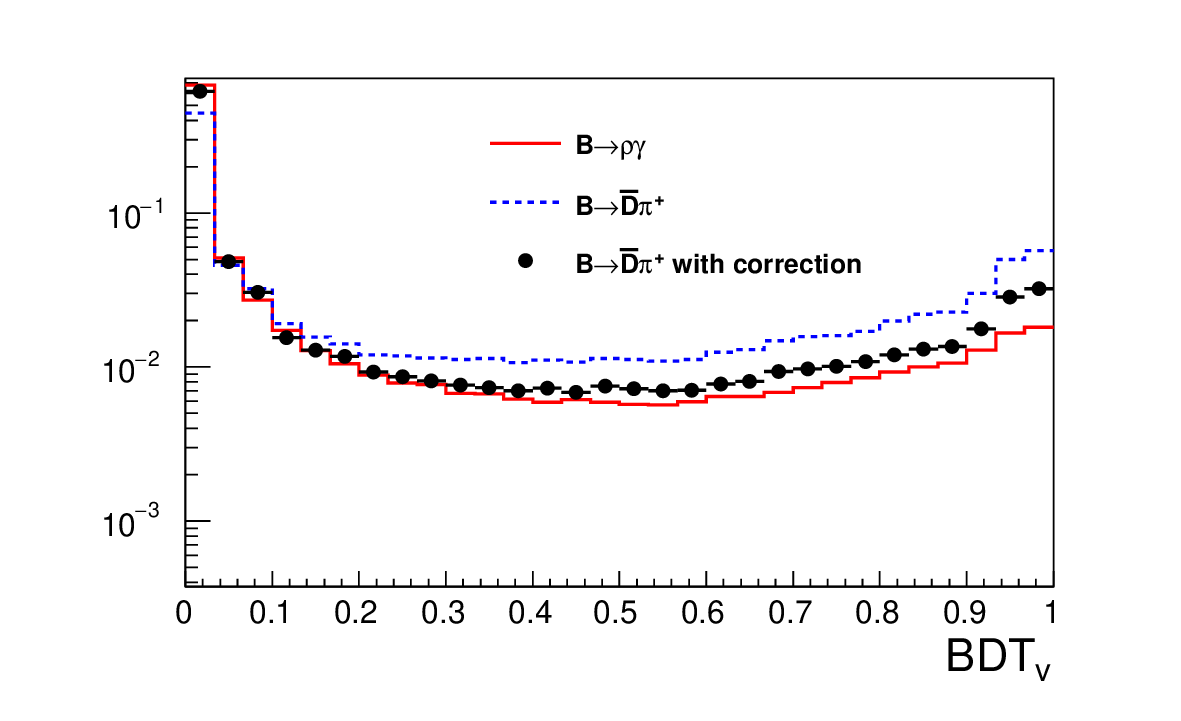}
 \caption{Distributions of $BDT_{v}$ for simulated data, for Belle (top) and Belle~II (bottom). The solid red histograms are $B\rightarrow\rho\gamma$, the dotted blue histograms are $B\rightarrow\overline{D}\pi^{+}$ and the points are the $B\rightarrow\overline{D}\pi^{+}$ with $M(\pi\gamma_{\rm soft})$ correction.}
 \label{fig:pi0prob_control}
\end{figure}

The signal efficiency is checked using a $B\rightarrow\overline{D}\pi^{+}$ control sample.
The $\pi^{+}$ momentum is scaled by a factor 1.087 to match the momentum of $\gamma_{\rm prim}$, and the mass $M(\pi\gamma_{\rm soft})$ is used in place of $M(\gamma_{\rm prim}\gamma_{\rm soft})$.
Due to isolation criteria applied to the $\gamma_{\rm prim}$ candidate, the distribution of $M(\pi\gamma_{\rm soft})$ differs from that of $M(\gamma_{\rm prim}\gamma_{\rm soft})$;
to correct for this, the $B\rightarrow\overline{D}\pi^{+}$ events are weighted such that these distributions match.
After this weighting, the $BDT_{v}$ distributions show improved agreement with MC-simulated $B\rightarrow\rho\gamma$ events, as shown in Fig.~\ref{fig:pi0prob_control}.
The difference in signal efficiency before and after this weighting is taken as a systematic uncertainty.

\subsection{$q\overline{q}$ suppression}

To reduce the remaining contamination from continuum background, we introduce another boosted decision tree classifier ($BDT_{q\overline{q}}$) that relies on differences in event shapes between $B\overline{B}$ and continuum $q\overline{q}$ events.
The variables used in $BDT_{q\overline{q}}$ are as follows:
the cosine of the angle between the thrust axis of the signal $B$ candidate and the thrust axis of the rest of the event $\cos{\theta_{T_{B}T_{O}}}$~\cite{Farhi:1977sg};
modified Fox-Wolfram moments~\cite{fox,pi0pi0};
the outputs of the $B$ flavor-tagging algorithms of Belle~\cite{Belle:2004uxp} and Belle~II~\cite{Belle-II:2021zvj};
the difference between the $z$ coordinates of the signal and the companion $B$ decay vertices;
the distance of closest approach between the signal $\pi^{+}$ track and the companion $B$ decay vertex;
the sphericity of non-signal particles in the event~\cite{sphericity};
and the cosine of the polar angle of the signal $B$.
The $\cos{\theta_{T_{B}T_{O}}}$ variable is the most discriminating.
The flavor-tagging algorithm outputs two quantities: the flavor tag $q$ ($=\pm 1$), and a quality factor $r$ that ranges from zero for no flavor discrimination to one for unambiguous flavor assignment.
The $BDT_{q\overline{q}}$ training uses simulated $B\rightarrow\rho\gamma$ and continuum events.
The training is performed separately for $B^{+}\rightarrow\rho^{+}\gamma$ and $B^{0}\rightarrow\rho^{0}\gamma$ decays, and for Belle and Belle~II data.
The requirements on the $BDT_{q\overline{q}}$ output are chosen to maximize the signal significance, separately for three ranges of the flavor-tagging quality $r$:
$(0,0.4)$, $(0.4,0.825)$, and $(0.825,1)$.
All requirements are greater than 0.95.
The signal efficiency is estimated from simulation and
corrected for possible differences between data and simulation using
the ratio of efficiencies between data and simulation for the $B\rightarrow K^{\ast}\gamma$ control samples.
These correction factors range from 0.96 to 1.10.
The signal efficiencies in Belle data for the $BDT_{q\overline{q}}$ requirement are 30\% and 40\% for $B^{+}$ and $B^{0}$ decays, respectively.
For Belle~II data, the corresponding efficiencies are 45\% and 52\%.


\section{Signal extraction}
\label{sec:fit}
To measure the $B\rightarrow\rho\gamma$ branching fractions and asymmetries $A_{\rm I}$ and $A_{\it CP}$, we perform an extended unbinned maximum likelihood fit to the observables $M_{\rm bc}$, $\Delta E$, and $M(K\pi)$.
We simultaneously fit six independent data sets:
$B^{+}$, $B^{-}$, and $(B^{0}+\overline{B}^{0})$ candidates combined, for Belle and Belle~II data.
In the fit, we model four components: signal, $B\rightarrow K^{\ast}\gamma$, $B\overline{B}$ background, and $q\overline{q}$ background.
The probability density functions (PDFs) used for these components are as follows.

For signal decays, $M_{\rm bc}$, $\Delta E$, and $M(K\pi)$ are found to be essentially uncorrelated and the PDF is taken to be the product of separate functions for each.
We model both the $M_{\rm bc}$ and $\Delta E$ distributions with Crystal Ball functions~\cite{Oreglia:1981fx},
and the $M(K\pi)$ distribution with a Novosibirsk function~\cite{Belle:1999bhb} convolved with a Gaussian.
For $B\rightarrow K^{\ast}\gamma$ background, $M_{\rm bc}$, $\Delta E$, and $M(K\pi)$ are correlated,
and we use a three-dimensional histogram PDF.
For $B\overline{B}$ background, we again use the product of three one-dimensional functions:
the sum of a Crystal Ball function and ARGUS function~\cite{ARGUS:1990hfq} for $M_{\rm bc}$,
an exponential function for $\Delta E$, and a histogram PDF for $M(K\pi)$.
Finally, for $q\overline{q}$ background, we use the product of an ARGUS function for $M_{\rm bc}$, a quadratic Chebychev polynomial for $\Delta E$,
and a Novosibirsk function convolved with a Gaussian for $M(K\pi)$.

The shape parameters of the $\Delta E$ distribution for $q\overline{q}$ background are floated in the fit,
while all other PDF shape parameters are fixed to values obtained from MC simulation.
To account for possible differences between data and simulation,
the $M_{\rm bc}$ and $\Delta E$ distributions for signal decays are corrected according to small differences observed between data and simulation for the $B\rightarrow K^{\ast}\gamma$ control samples.
Similarly, the parameters for the $M_{\rm bc}$ and $M(K\pi)$ distributions of $q\overline{q}$ background are corrected according to small differences observed between data and simulation in the sideband region $BDT_{q\overline{q}}\in [0.6,0.9]$.
In addition to these fixed shapes, we also fix the yields of $B\rightarrow K^{\ast}\gamma$ and $B\overline{B}$ backgrounds to expectations based on MC simulation.

With the above PDFs, we perform two fits.
We first fit directly for parameters $D$, $A_{\rm I}$, and $A_{\it CP}$, defined as
\begin{linenomath}
 \begin{align}
  & D = c_{\rho}^{2}\frac{\tau_{B^{\pm}}}{\tau_{B^{0}}}\mathcal{B}(\overset{\textbf{\fontsize{2pt}{2pt}\selectfont(---)}}{B^{0}}\rightarrow\rho^{0}\gamma)+\mathcal{B}\left(B^{\pm}\rightarrow\rho^{\pm}\gamma\right),\label{eq:D}\\
  & A_{\rm I}=\frac{ c_{\rho}^{2}\frac{\tau_{B^{\pm}}}{\tau_{B^{0}}}\mathcal{B}(\overset{\textbf{\fontsize{2pt}{2pt}\selectfont(---)}}{B^{0}}\rightarrow\rho^{0}\gamma)-\mathcal{B}\left(B^{\pm}\rightarrow\rho^{\pm}\gamma\right) }{D},\label{eq:ai}\\
  \nonumber
  {\rm and}\\
  & A_{\it CP}=\frac{ N_{B^{-}} - N_{B^{+}} }{ N_{B^{-}} + N_{B^{+}} },\label{eq:acp}
 \end{align}
\end{linenomath}

where $\tau_{B^\pm}/\tau_{B^0} = 1.076\pm0.004$~\cite{ParticleDataGroup:2022pth}
is the ratio of $B^\pm$ to $B^0$ lifetimes.

\begin{figure*}[h]
 \includegraphics[width=80mm]{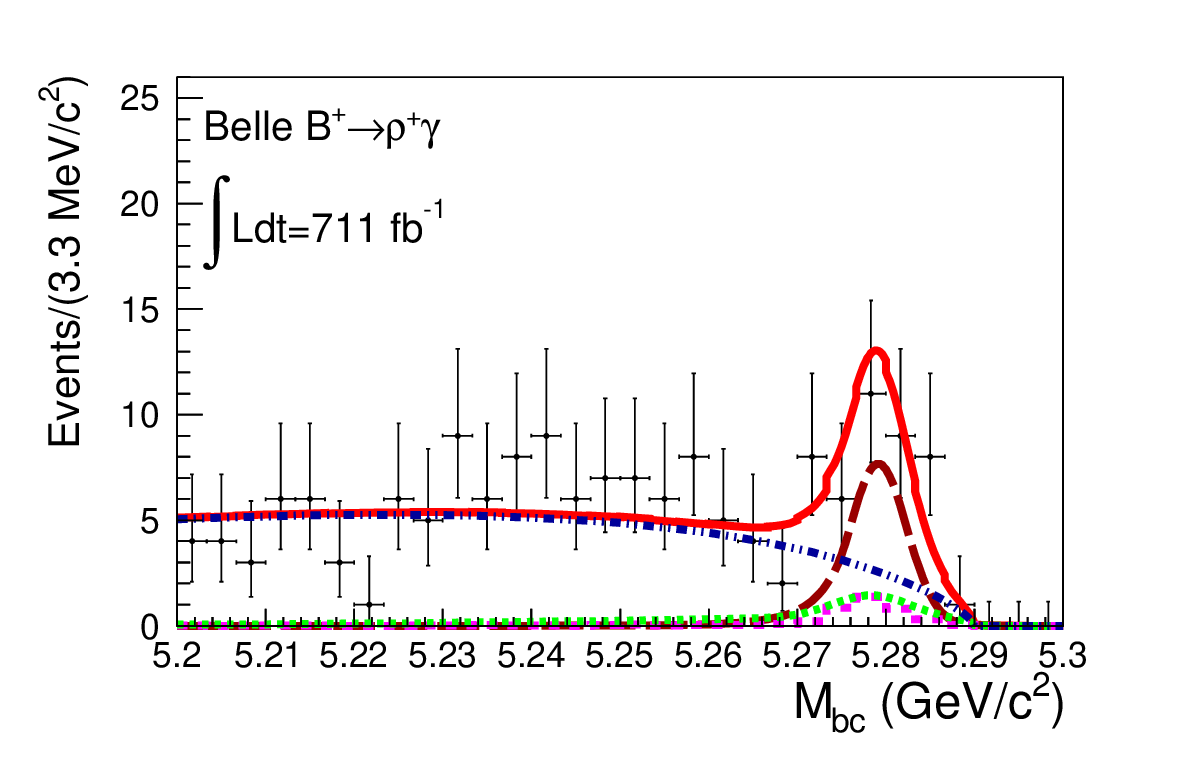}
 \includegraphics[width=80mm]{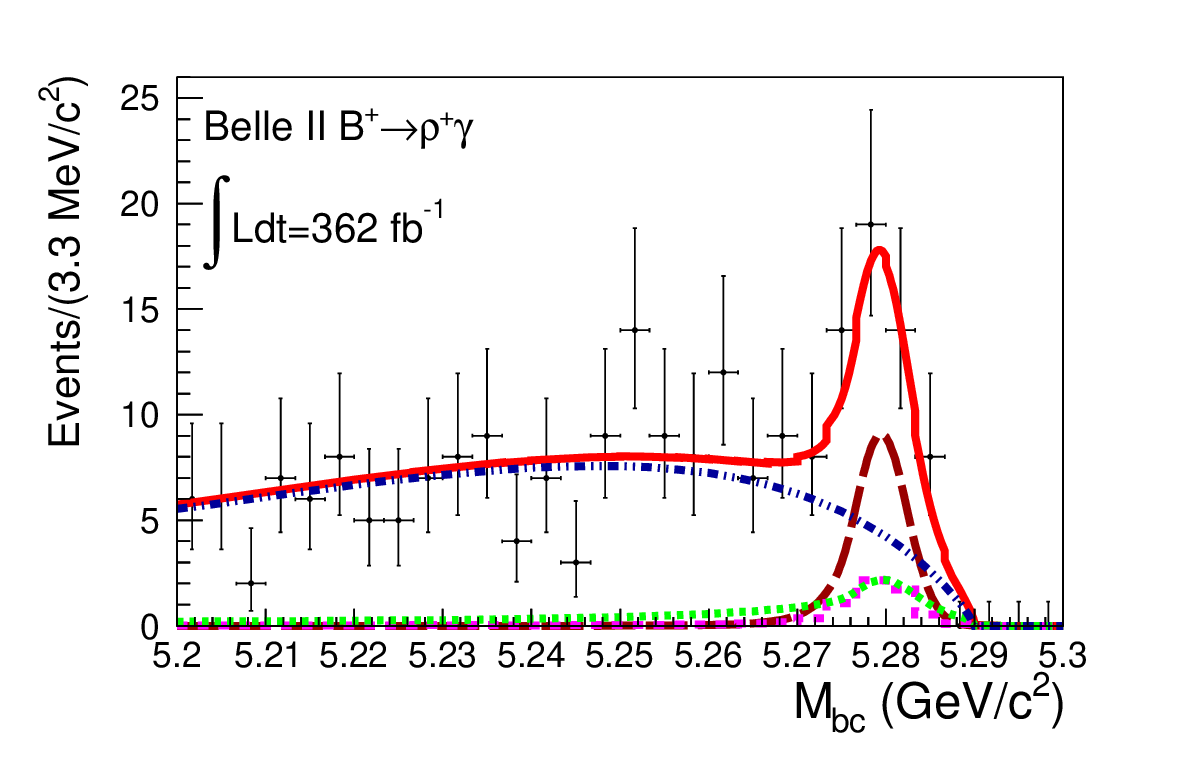}\\
 \includegraphics[width=80mm]{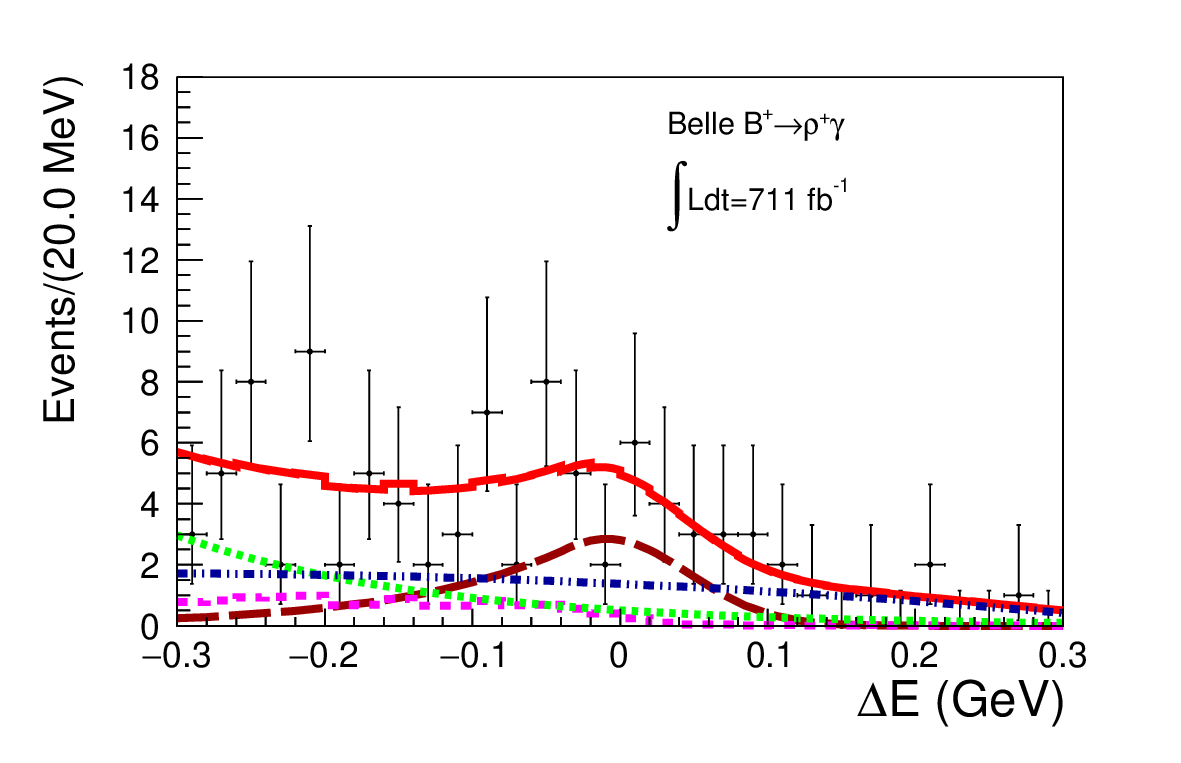}
 \includegraphics[width=80mm]{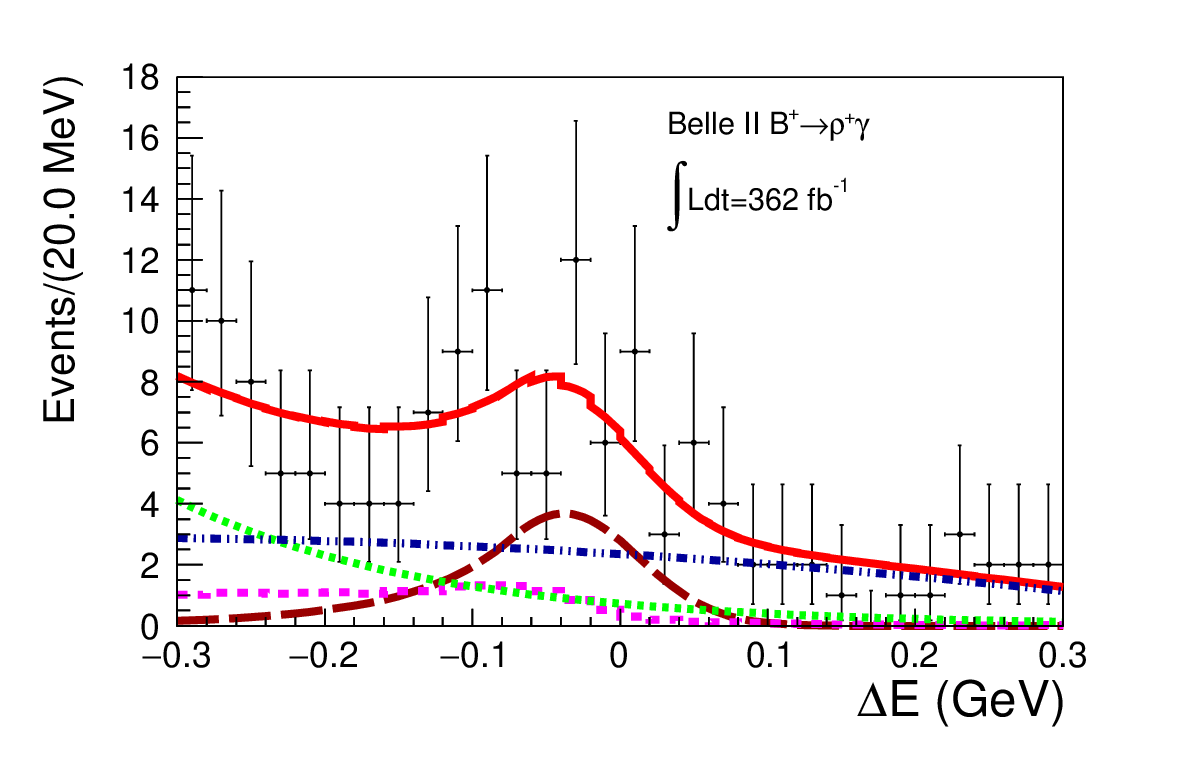}\\
 \includegraphics[width=80mm]{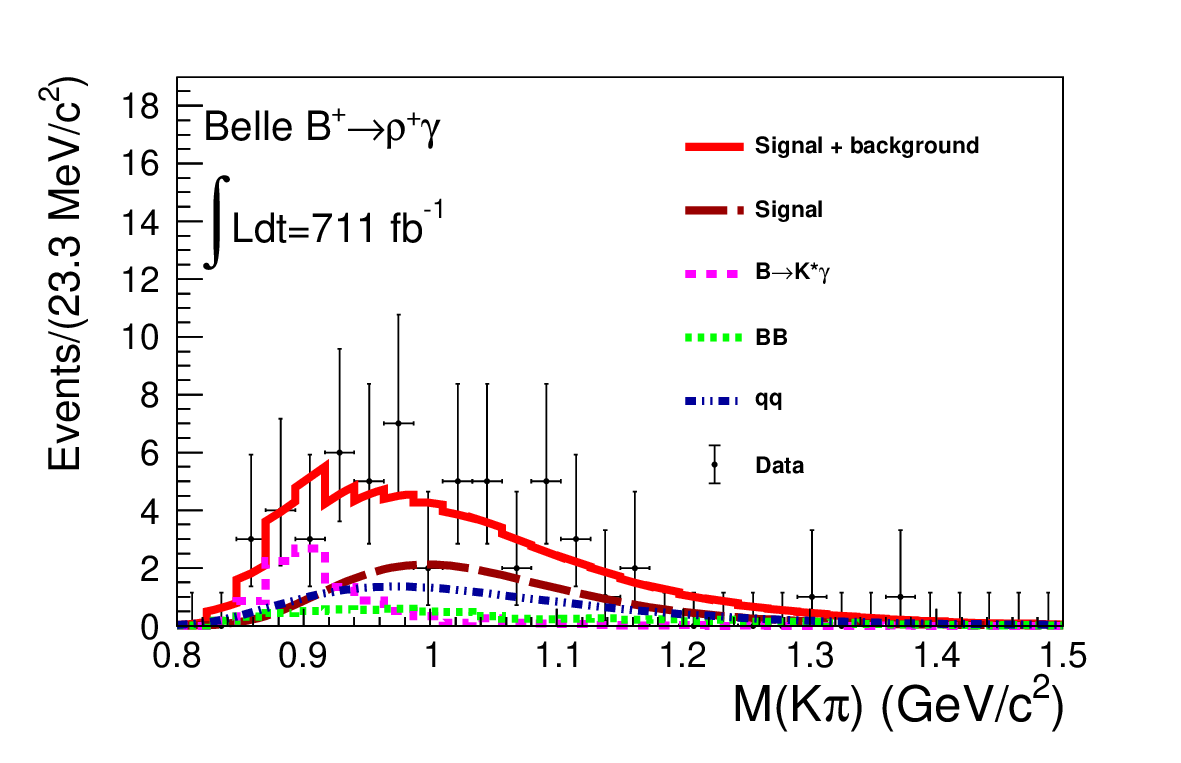}
 \includegraphics[width=80mm]{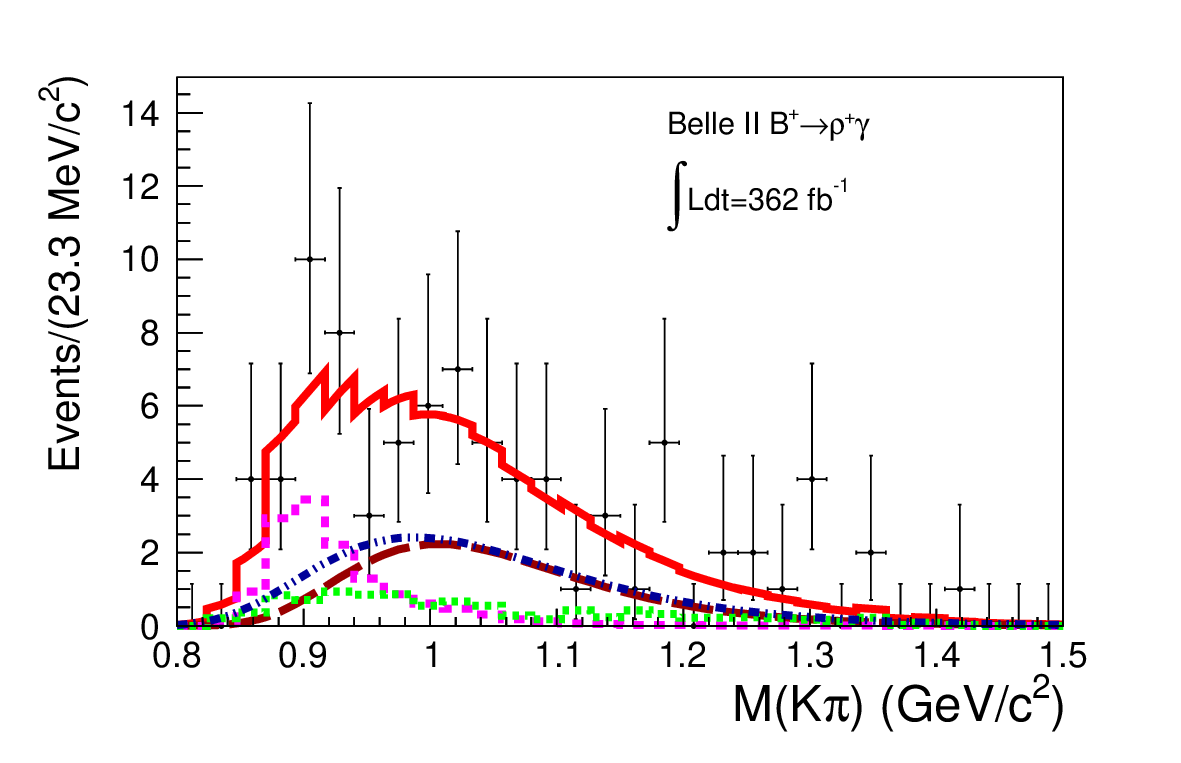}
 \caption{Distributions of $M_{\rm bc}$, $\Delta E$ and $M(K\pi)$ for $B^{+}\rightarrow\rho^{+}\gamma$ candidates reconstructed in the signal-enriched region of the other two variables. The signal-enriched region is defined as $M_{\rm bc}>5.27~{\mathrm {GeV}}/c^{2}$, $|\Delta E|<0.1~{\mathrm {GeV}}$ and $M(K\pi)>0.92~{\mathrm {GeV}}/c^{2}$. The points with error bars are data, the solid red curves are the sum of signal and background PDFs, the dashed red curves are signal, the dotted-dashed blue curves are continuum background, the dashed magenta curves are $K^{\ast}\gamma$ background, and the dotted green curves are $B\overline{B}$ background other than $K^{\ast}\gamma$. The discrete nature of the solid red curves is due to the use of histogram PDFs.}
 \label{fig:fit1}
\end{figure*}

\begin{figure*}[h]
 \includegraphics[width=80mm]{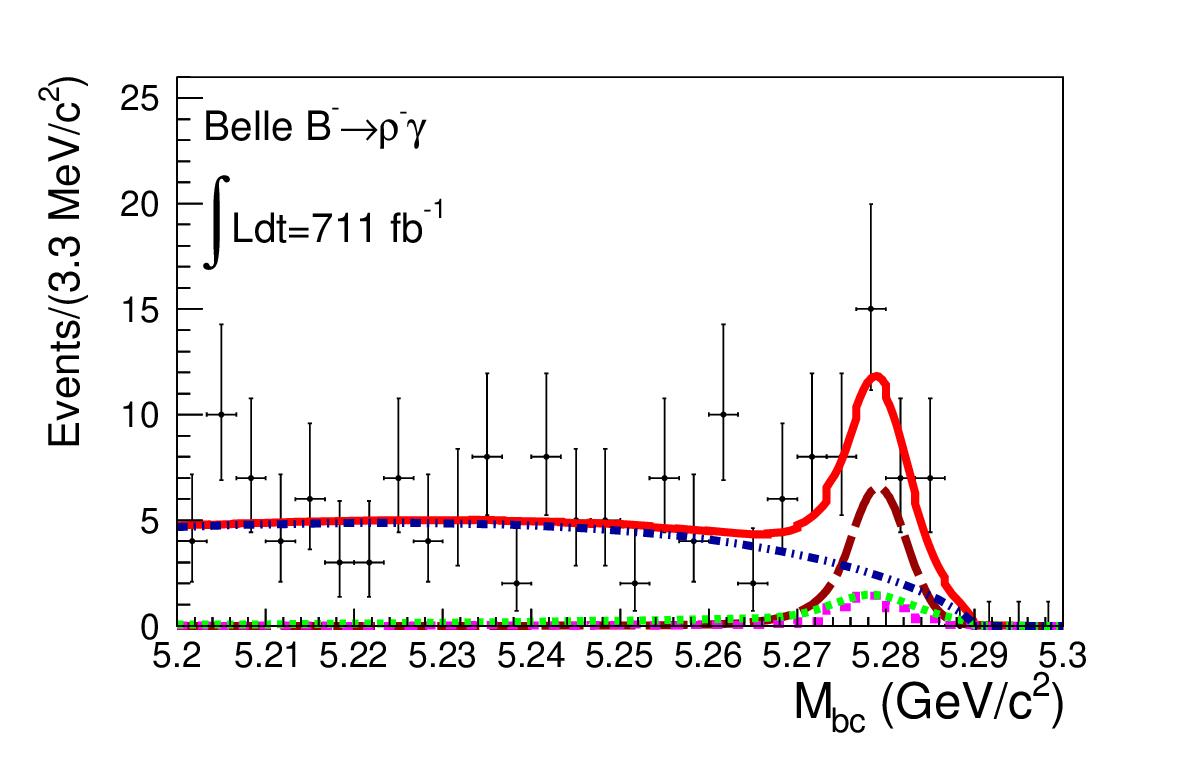}
 \includegraphics[width=80mm]{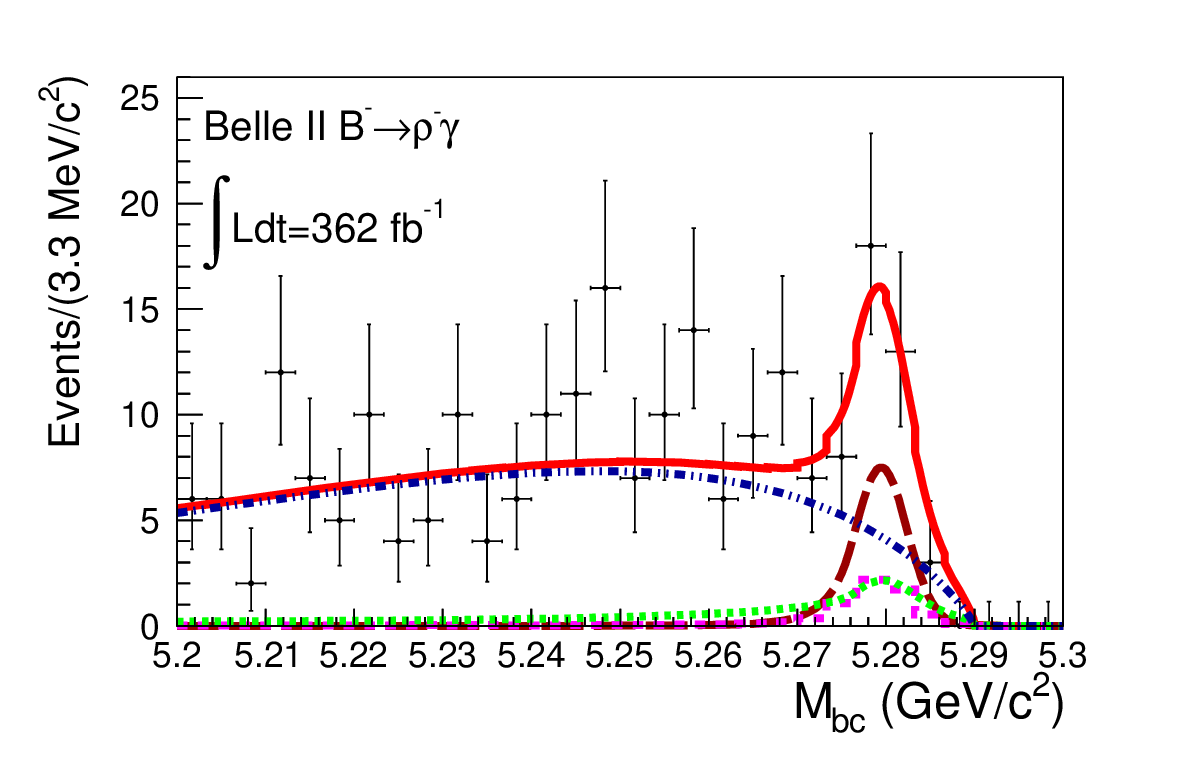}\\
 \includegraphics[width=80mm]{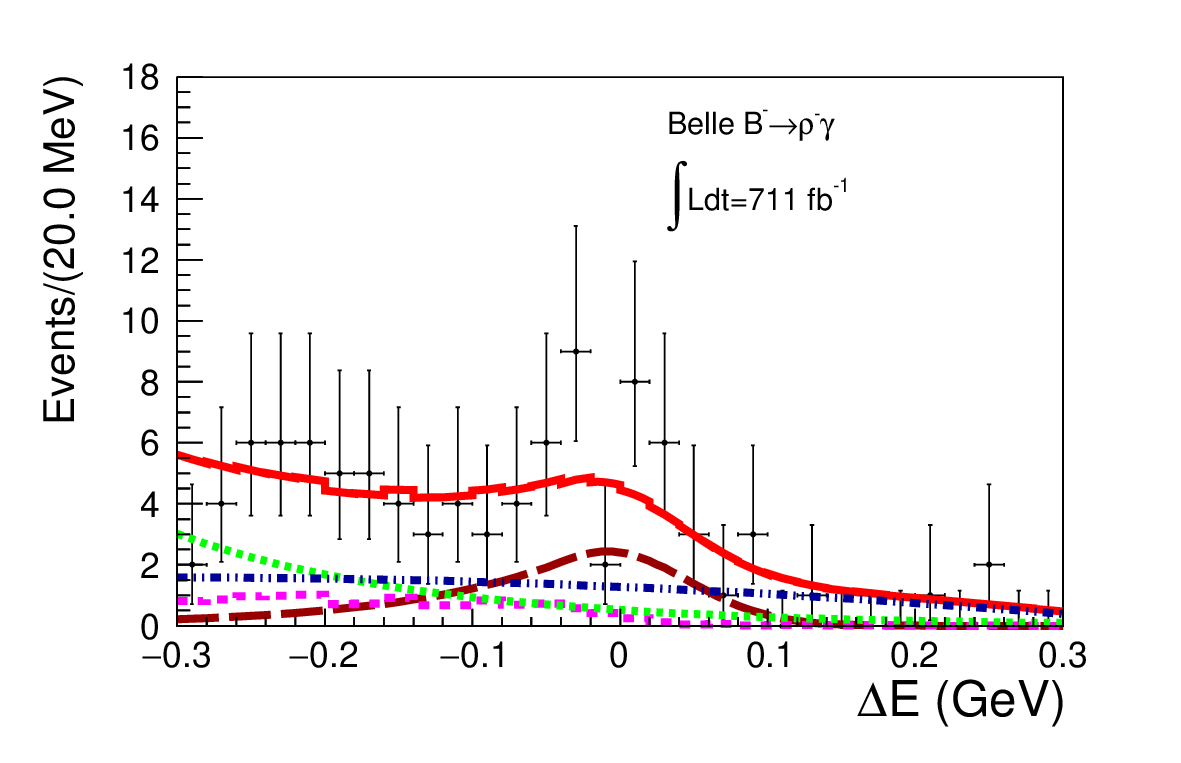}
 \includegraphics[width=80mm]{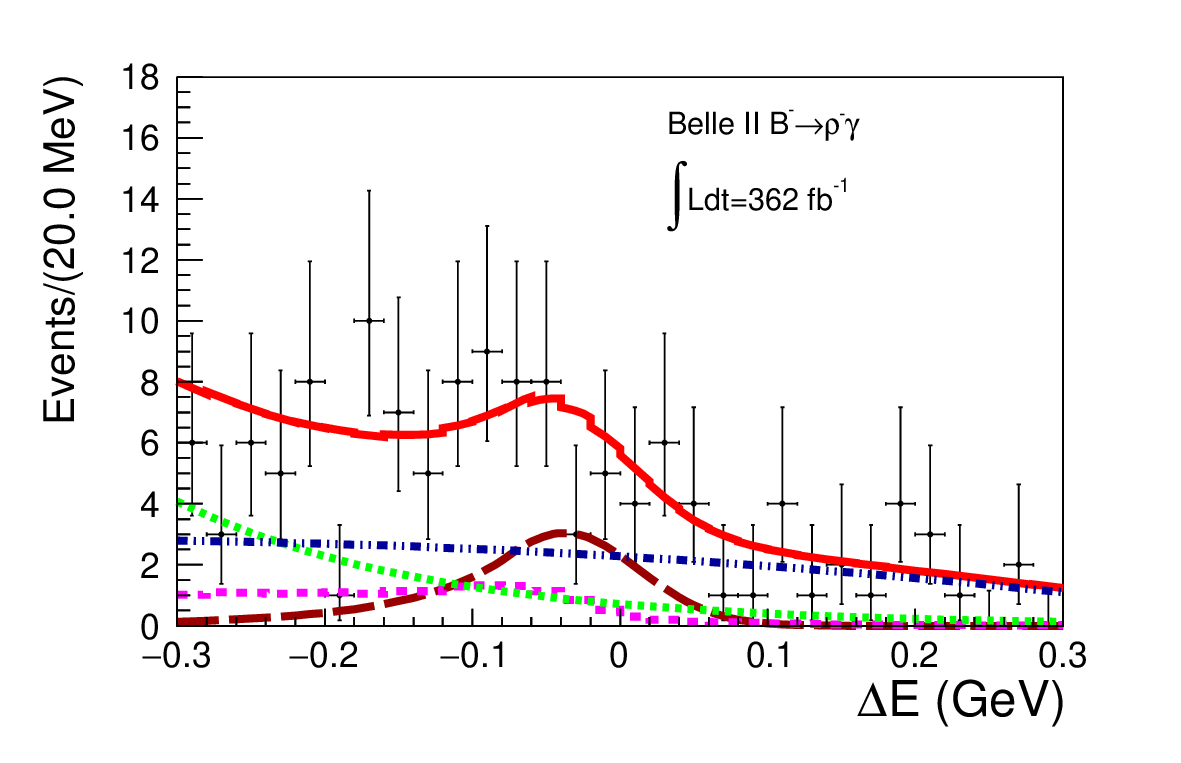}\\
 \includegraphics[width=80mm]{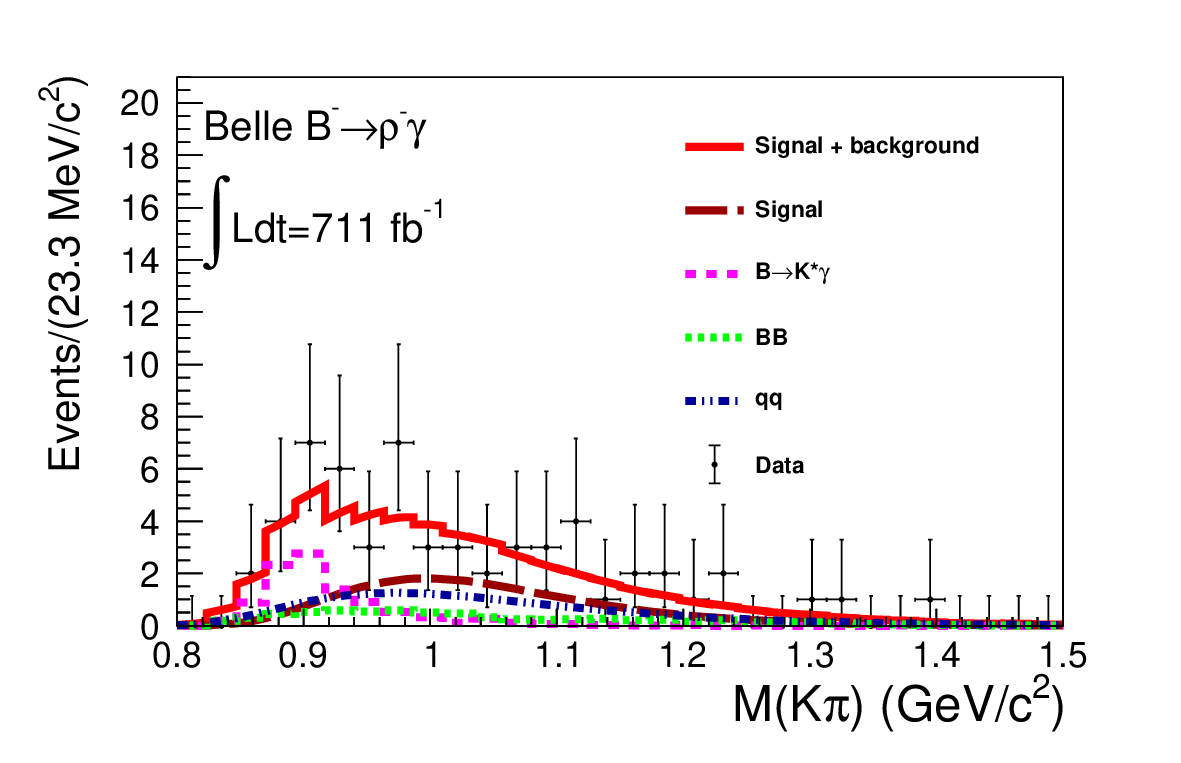}
 \includegraphics[width=80mm]{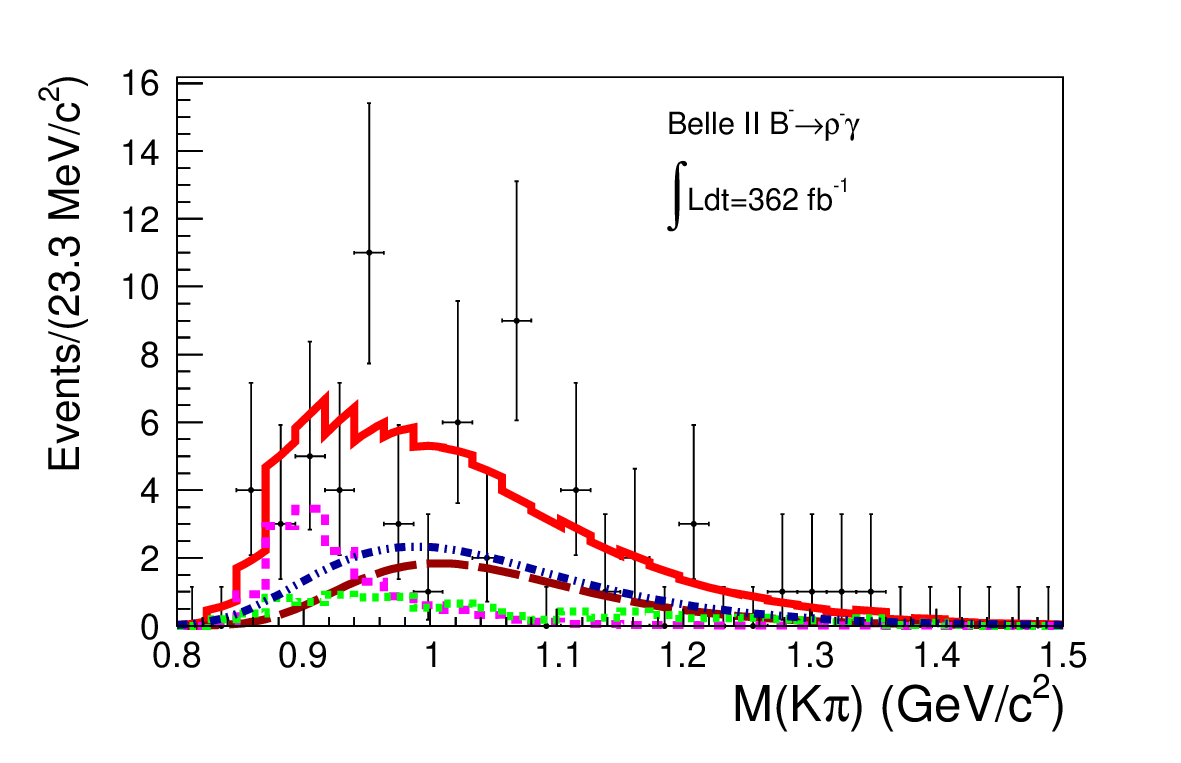}
 \caption{Distributions of $M_{\rm bc}$, $\Delta E$ and $M(K\pi)$ for $B^{+}\rightarrow\rho^{+}\gamma$ candidates reconstructed in the signal-enriched region of the other two variables. The signal-enriched region is defined as $M_{\rm bc}>5.27~{\mathrm {GeV}}/c^{2}$, $|\Delta E|<0.1~{\mathrm {GeV}}$ and $M(K\pi)>0.92~{\mathrm {GeV}}/c^{2}$. The points with error bars are data, the solid red curves are the sum of signal and background PDFs, the dashed red curves are signal, the dotted-dashed blue curves are continuum background, the dashed magenta curves are $K^{\ast}\gamma$ background, and the dotted green curves are $B\overline{B}$ background other than $K^{\ast}\gamma$. The discrete nature of the solid red curves is due to the use of histogram PDFs.}
 \label{fig:fit2}
\end{figure*}

\begin{figure*}[h]
 \includegraphics[width=80mm]{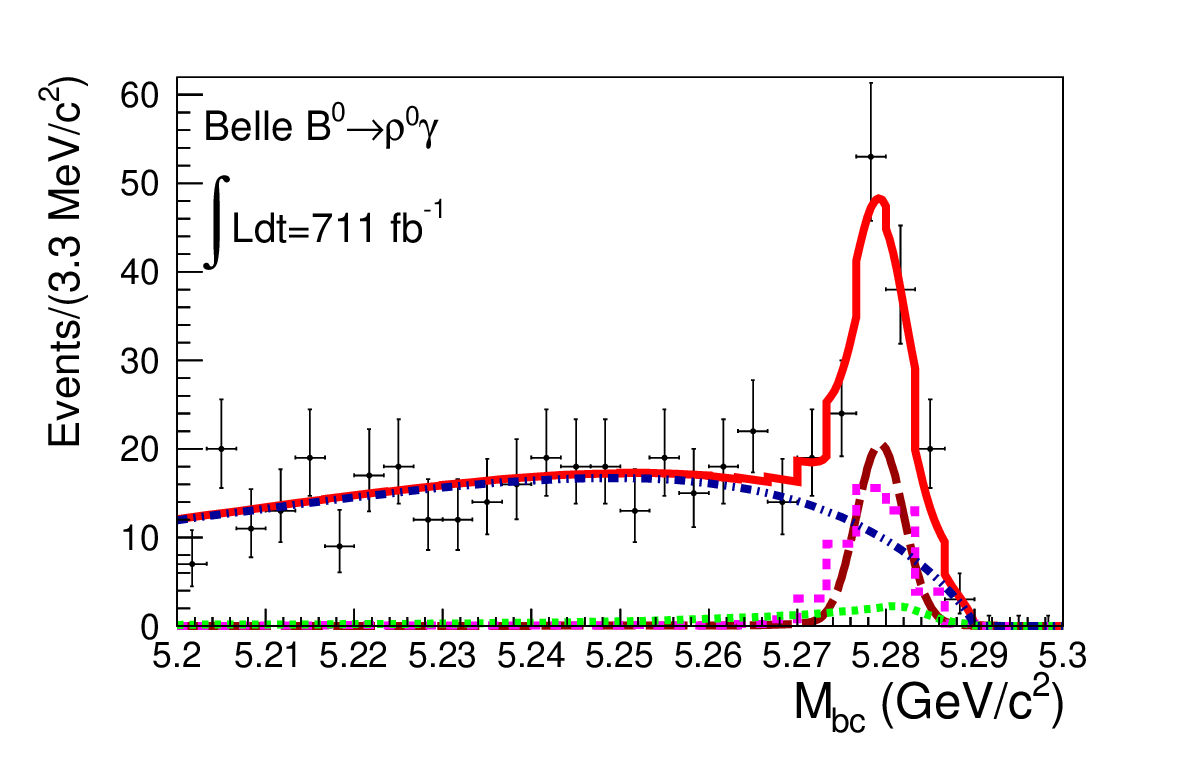}
 \includegraphics[width=80mm]{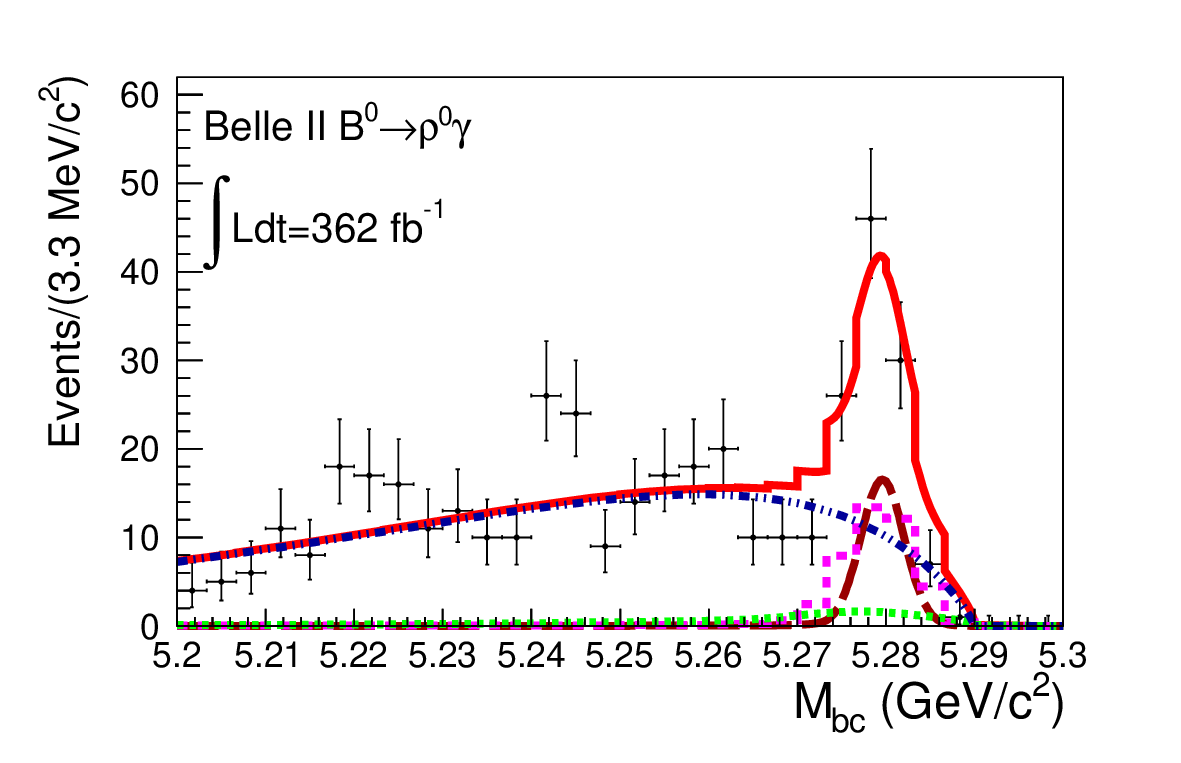}\\
 \includegraphics[width=80mm]{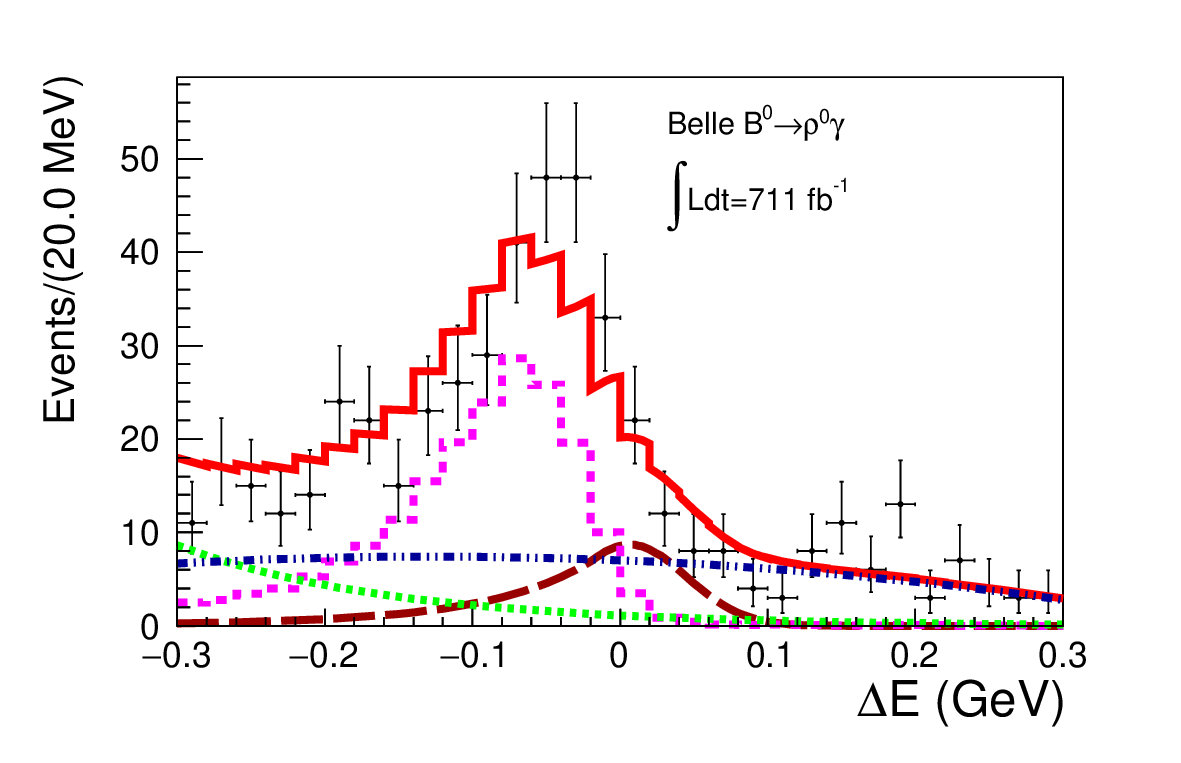}
 \includegraphics[width=80mm]{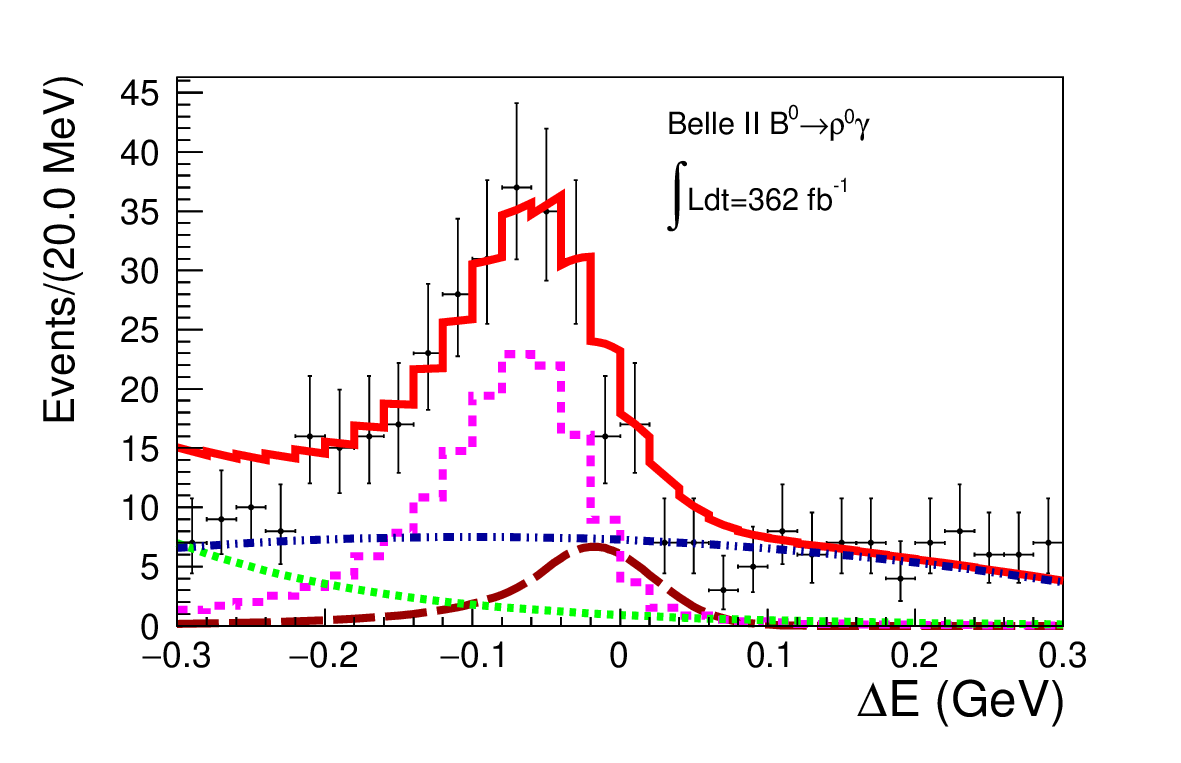}\\
 \includegraphics[width=80mm]{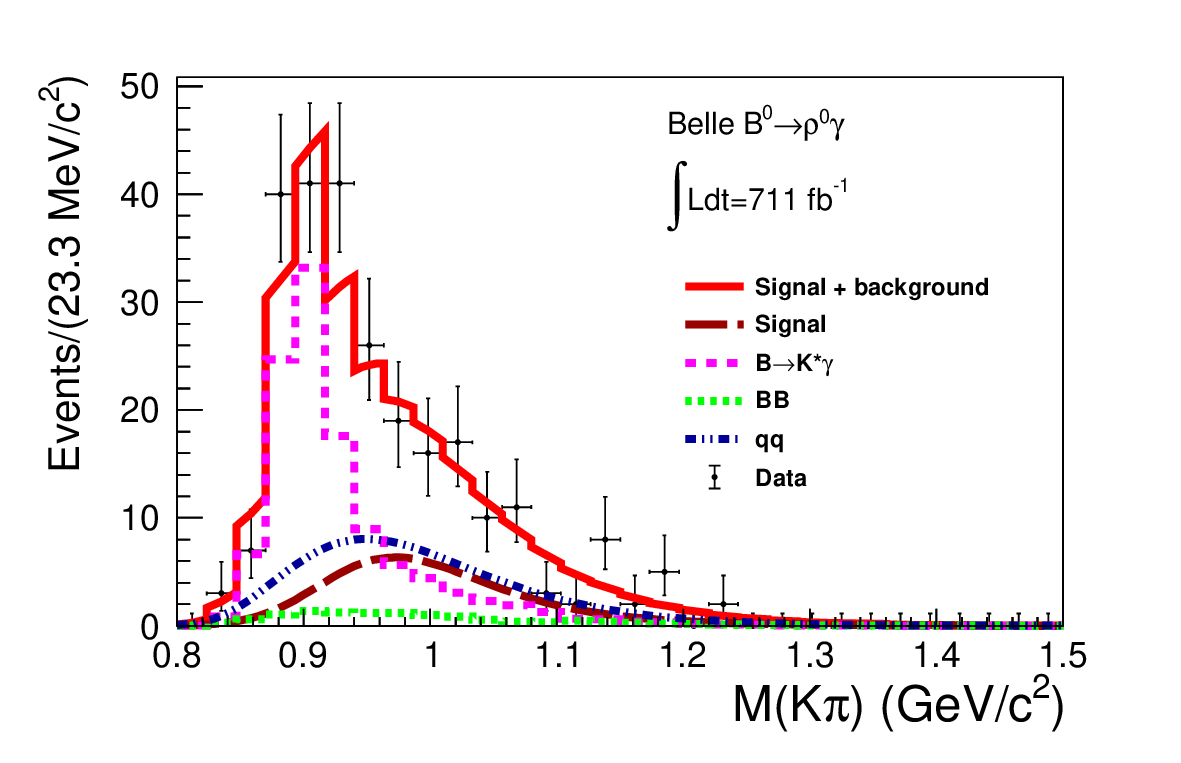}
 \includegraphics[width=80mm]{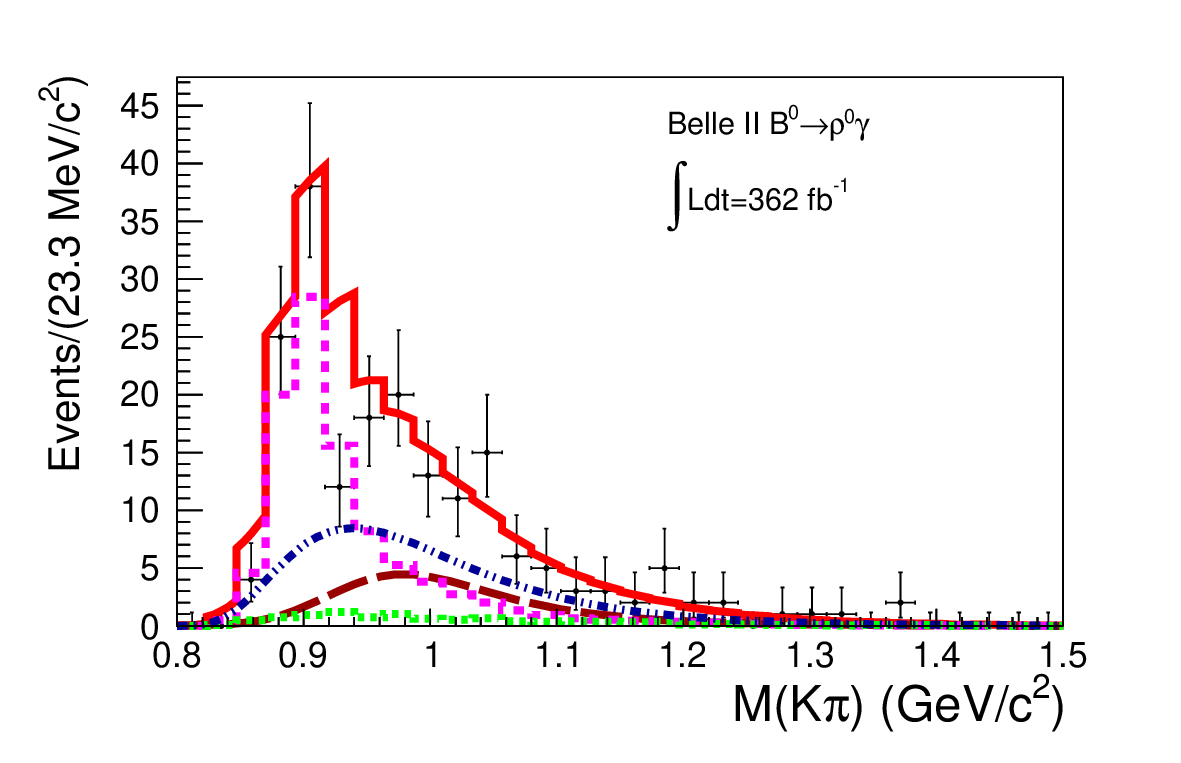}
 \caption{Distributions of $M_{\rm bc}$, $\Delta E$ and $M(K\pi)$ for $B^{+}\rightarrow\rho^{+}\gamma$ candidates reconstructed in the signal-enriched region of the other two variables. The signal-enriched region is defined as $M_{\rm bc}>5.27~{\mathrm {GeV}}/c^{2}$, $|\Delta E|<0.1~{\mathrm {GeV}}$ and $M(K\pi)>0.92~{\mathrm {GeV}}/c^{2}$. The points with error bars are data, the solid red curves are the sum of signal and background PDFs, the dashed red curves are signal, the dotted-dashed blue curves are continuum background, the dashed magenta curves are $K^{\ast}\gamma$ background, and the dotted green curves are $B\overline{B}$ background other than $K^{\ast}\gamma$. The discrete nature of the solid red curves is due to the use of histogram PDFs.}
 \label{fig:fit3}
\end{figure*}

The relationship between the number of signal events ($N_{i}$)
and the branching fractions in Eq.~\ref{eq:ai} and \ref{eq:acp} 
is given by
\begin{linenomath}
 \begin{equation}
  {\cal B}(B^{\pm (0)}\rightarrow\rho^{\pm (0)}\gamma)=\frac{N_{B^{-}(\overline{B}{}^{0})}+N_{B^{+}(B^{0})}}{2N_{B\overline{B}}f_{+-(00)}\epsilon^{\pm (0)}}.\label{eq:br}
 \end{equation}
\end{linenomath}
Here, 
$N_{B\overline{B}}$ is the number of $B\overline{B}$ pairs and equals 
$(771.6\pm 10.6)\times 10^{6}$ for Belle and
$(378.5\pm 5.8)\times 10^{6}$ for Belle~II;
$f_{+-}/f_{00}=1.065\pm 0.052$~\cite{Belle:2022hka}
\footnote{For $a=f_{+-}/f_{00}$, $f_{+-}=a/(1+a)$ and $f_{00}=1/(1+a)$.}
is the production ratio of $B^+B^-$ to $B^0\overline{B}{}^{0}$ pairs at an $e^+e^-$ center-of-mass energy
corresponding to the $\Upsilon (4S)$ resonance;
and $\epsilon^{\pm}$ and $\epsilon^{0}$ are reconstruction efficiencies for $B^{\pm}\rightarrow\rho^{\pm}\gamma$ and $B^{0}\rightarrow\rho^{0}\gamma$, respectively.
The efficiencies of $B^{+}\rightarrow\rho^{+}\gamma$ and $B^{-}\rightarrow\rho^{-}\gamma$ decays are consistent with each other and taken to be identical.
The signal efficiencies $\epsilon^{\pm}$ and $\epsilon^{0}$ for the Belle (Belle~II) sample are 5.5\% and 10.3\% (11.0\% and 14.9\%), respectively.
The higher efficiency for Belle~II is due 
mainly to the improved performance of the $BDT_{v}$ and $BDT_{q\overline{q}}$ algorithms.
After this first fit, we perform a second fit directly for 
$\mathcal{B}\left(B^{+}\rightarrow\rho^{+}\gamma\right)$ and 
$\mathcal{B}\left(B^{-}\rightarrow\rho^{-}\gamma\right)$
to obtain their uncertainties
(rather than unfolding them from the first fit results, 
accounting for correlations).
The distributions of $M_{\rm bc}$, $\Delta E$, and $M(K\pi)$ are shown in Figs.~\ref{fig:fit1}--\ref{fig:fit3} along with projections of the fit result.

Using Eqs.~\ref{eq:D}-\ref{eq:br}, we extract the signal yields; 
these are listed in Table~\ref{tb:nsig} along with the corresponding reconstruction efficiencies and the fitted yields of $q\overline{q}$ background.
\begin{table}[h]
 \begin{center}
  \caption{Efficiency ($\epsilon$), signal yield ($N_{\rho\gamma}$) and continuum background yield ($N_{q\overline{q}}$) from Belle (711~${\rm fb^{-1}}$) and Belle~II (362~${\rm fb^{-1}}$) data. The uncertainties for $N_{\rho\gamma(q\overline{q})}$ are statistical only.}
  \label{tb:nsig}
  \begin{tabular}{l c c c}\hline\hline
   Mode & $\epsilon~[\%]$ & $N_{\rho\gamma}$ & $N_{q\overline{q}}$ \\ \hline\hline
   Belle $B^{+}\rightarrow\rho^{+}\gamma$ & $5.5\pm 0.5$ & $31\pm 6$ & $400\pm 21$ \\
   Belle $B^{-}\rightarrow\rho^{-}\gamma$ & $5.5\pm 0.5$ & $26\pm 6$ & $369\pm 21$ \\
   Belle $B^{0}\rightarrow\rho^{0}\gamma$ & $10.3\pm 0.4$ & $58\pm 10$ & $1307\pm 38$ \\
   Belle~II $B^{+}\rightarrow\rho^{+}\gamma$ & $11.0\pm 1.1$ & $30\pm 6$ & $552\pm 25$ \\
   Belle~II $B^{-}\rightarrow\rho^{-}\gamma$ & $11.0\pm 1.1$ & $26\pm 6$ & $549\pm 25$ \\
   Belle~II $B^{0}\rightarrow\rho^{0}\gamma$ & $14.9\pm 0.5$ & $41\pm 7$ & $1114\pm 35$ \\ \hline
  \end{tabular}
 \end{center}
\end{table}
The overall signal yields in the full fitted region are $114\pm 12$ $B^{\pm}\rightarrow\rho^{\pm}\gamma$ decays and $99\pm 12$ $B^{0}\rightarrow\rho^{0}\gamma$ decays.
As a cross-check, we fit for the branching fractions using only Belle data;
our results are in agreement with the previous Belle measurement~\cite{Belle:2008imr}.

\section{Systematic uncertainties}
\label{sec:systematics}

The various sources of systematic uncertainty are listed 
in Table~\ref{tb:syst} and estimated as follows.


The systematic uncertainty arising from particle
reconstruction consists of several parts.
The uncertainty related to photon detection
is measured in Belle using  $e^+e^-\rightarrow e^+e^-\gamma$ 
(radiative Bhabha) events and in Belle~II
using $e^{+}e^{-}\rightarrow\mu^{+}\mu^{-}\gamma$ events;
the resulting uncertainties are 2\% and 1\%, respectively.
The uncertainty due to $\pi^{+}$ identification is measured 
in Belle (0.7\%) and Belle~II (0.3\%) using a sample of
$D^{\ast +}\rightarrow D^{0}(\rightarrow K^-\pi^+)\pi^+$ decays.
The uncertainty due to track reconstruction is evaluated in 
Belle (0.3\%) using
$D^{\ast +}\to D^0 (\to K_S\,\pi^+\pi^-) \pi^+$ decays,
and in Belle~II (0.3\%) using
$e^{+}e^{-}\rightarrow\tau^{+}\tau^{-}$ events.
The uncertainty due to $\pi^{0}$ reconstruction is estimated by comparing
the ratio of efficiencies for two decay channels with different numbers 
of $\pi^{0}$'s, for both data and simulation.
For Belle, the ratio is the efficiency for 
$\eta\rightarrow\pi^{0}\pi^{0}\pi^{0}$ to that for 
$\eta\rightarrow\pi^{+}\pi^{-}\pi^{0}$ or $\eta\rightarrow\gamma\gamma$;
the resulting uncertainty is 3\%.
For Belle~II the ratio is the efficiency for 
$D^{0}\rightarrow K^{-}\pi^{+}\pi^{0}$ to that for 
$D^{0}\rightarrow K^{-}\pi^{+}$, and the resulting
uncertainty is 5\%.

\begin{table}[hbt]
\begin{center}
\caption{Systematic uncertainties on the branching fractions for $B^{+}\rightarrow\rho^{+}\gamma$ (${\cal B}_{\rho^{+}\gamma}$) and $B^{0}\rightarrow\rho^{0}\gamma$ (${\cal B}_{\rho^{0}\gamma}$), and on the isospin and {\it CP} asymmetries.}
\label{tb:syst}
\begin{tabular}{l c c c c}\hline\hline
 Source & ${\cal B}_{\rho^{+}\gamma}\times 10^{8}$ & ${\cal B}_{\rho^{0}\gamma}\times 10^{8}$ & $A_{\rm I}$ & $A_{\it CP}$ \\ \hline\hline
 Particle detection                   & 4.1 & 1.3 & 1.4\% & 0.5\% \\
 Selection criteria & 9.0 & 3.4 & 4.0\% & 0.5\% \\
 Fixed fit parameters                 & 1.1 & 2.7 & 1.8\% & 0.2\% \\
 Signal shape                         & 4.7 & 3.0 & 3.1\% & 0.5\% \\
 Histogram PDFs                       & 1.0 & 0.6 & 0.5\% & 0.1\% \\
 Peaking $K^{\ast}\gamma$ bkg         & 3.4 & 5.4 & 3.1\% & 0.1\% \\
 Other peaking $B\overline{B}$ bkgs   & 2.2 & 0.8 & 0.9\% & 0.2\% \\
 Peaking $B\overline{B}$ $A_{\it CP}$ & 0.1 & $<$0.1 & 0.1\% & 1.0\% \\
 Number of $B\overline{B}$'s          & 1.7 & 1.4 & 0.3\% & 0.1\% \\
 $\tau_{B^{\pm}}/\tau_{B^{0}}$        & 0.1 & $<$0.1 & 0.2\% & $<$0.1\% \\
 $f_{+-}/f_{00}$                      & 4.0 & 3.6 & 3.8\% & $<$0.1\% \\ \hline\hline
 Total                                & 12.5 & 8.6 & 7.5\% & 1.4\% \\ \hline\hline
\end{tabular}
\end{center}
\end{table}

Systematic uncertainties due to selection criteria are 
evaluated as follows.
The uncertainty due to differences between data and simulation for the 
BDT requirements is evaluated by varying the calibration factors obtained 
from the $B\rightarrow \overline{D}\pi$ and $B\rightarrow K^{\ast}\gamma$ 
control samples by one standard deviation.
The resulting uncertainties arising from the $\pi^{0}$ veto 
are 0.8\% (0.7\%) for the $B^+\to\rho^+\gamma$ ($B^0\to\rho^0\gamma$)
branching fraction, and 0.7\% for the isospin asymmetry. 
The uncertainties arising from $q\overline{q}$ suppression 
along with the $\eta$ veto are 6.8\% (4.2\%) for the 
$B^+$ ($B^0$) branching fraction, and 3.9\% for~$A_I$.
The uncertainty due to possible mis-modeling of the $B\rightarrow\overline{D}\pi^{+}$ control sample in Belle (see section~\ref{sec:pi0etaveto}) is 0.9\% (1.5\%) for
the $B^+$ ($B^0$) branching fraction and 0.3\% for~$A_I$.
The uncertainties on $A_{CP}$ are~$O(0.1\%)$.


The uncertainty related to each fixed parameter in the fit is estimated by 
varying each parameter by its uncertainty and repeating the fit. The difference
between the fit result and our nominal result is taken as the systematic 
uncertainty. 



The uncertainty arising from the shape of the signal PDF is evaluated by
varying the calibration factors obtained from the fit to the
$B\rightarrow K^{\ast}\gamma$ control sample by their uncertainties
and repeating the fit. The difference between the result and our 
nominal result is taken as the systematic uncertainty. 


The uncertainty due to modeling
$B\rightarrow K^{\ast}\gamma$ is evaluated by using a modified histogram PDF; 
this is obtained from
simulated experiments generated with the original histogram as a kernel function.
The differences in the fit results from the nominal values are taken 
as the systematic uncertainties.
The same procedure is adopted to estimate the uncertainty due to the histogram PDF used to model the $M(K\pi)$ distribution of the $B\overline{B}$ combinatorial background.

The systematic uncertainties due to $B\overline{B}$ background that peaks in the signal-enriched region (``peaking background'') are estimated by varying this background as follows.
The $K^{\ast}\gamma$ background yield is estimated as 
$N_{K^{\ast}\gamma}=2\cdot N_{B\overline{B}}\cdot {\cal B}(B\to K^{\ast}\gamma)\cdot F_{K/\pi}\cdot\epsilon^{}_{K^{\ast}\gamma}$,
where $F_{K/\pi}$ is the probability for a charged kaon to be misidentified as a pion,
and $\epsilon^{}_{K^{\ast}\gamma}$ is the efficiency to select $K^{\ast}\gamma$ background.
The misidentification probability $F_{K/\pi}$ is obtained from a study of 
$D^{\ast +}\to D^0 (\to K^-\pi^+) \pi^+$ decays; the fractional uncertainty
is 6.1\% for Belle and 9.7\% for Belle~II. 
To estimate the systematic uncertainty arising from  
$B\rightarrow K^{\ast}\gamma$ background,
we vary both ${\cal B}(B\rightarrow K^{\ast}\gamma)$~\cite{ParticleDataGroup:2022pth} and $F_{K/\pi}$ by their uncertainties.
Other peaking background sources are $B\rightarrow X_{s(d)}\gamma$ decays,
where $X_{s(d)}$ is any final state of $s(d)$ quark hadronization with strangeness equal to one (zero) except for $K^{\ast}\left(892\right)\gamma$ ($\rho\gamma$),
and
hadronic $B$ decays with $\pi^{0}(\eta)\rightarrow\gamma\gamma$ in the final states.
The former is varied according to its experimental uncertainty~\cite{ParticleDataGroup:2022pth}.
For the latter, we take the fractional uncertainty on the number of peaking background events to be the weighted sum of the fractional uncertainties on the branching fractions for the main decay modes that contribute.
For the $B^{+}\rightarrow\rho^{+}\gamma$ mode, the contamination from $B^{+}\rightarrow\rho^{+}\pi^{0}(\eta)$ decays is dominant at 24\% (16\%), 
according to the simulation.
For the $B^{0}\rightarrow\rho^{0}\gamma$ mode, the main contamination comes from $B^{0}\rightarrow\rho^{0}\eta$ (9\%), $B^{+}\rightarrow a_{1}^{+}\pi^{0}$ (8\%), 
and $B^{+}\rightarrow\rho^{+}\rho^{0}$ (8\%).
For modes that have not been measured, a 50\% uncertainty is assumed 
for the branching fractions.

For the $A_{\it CP}$ measurement, we calculate the effect of a possible {\it CP} asymmetry of $B\overline{B}$ peaking background.
The $A_{\it CP}$ values for $B\rightarrow K^{\ast}\gamma$ and $B\rightarrow X_{s}\gamma$ are varied by their measured uncertainties~\cite{ParticleDataGroup:2022pth},
while $A_{CP}$ for $B\rightarrow X_{d}\gamma$ is varied by $\pm 60\%$.
The latter corresponds to a conservative uncertainty on the SM 
prediction~\cite{Benzke:2010tq}.
The $A_{\it CP}$ values for other modes such as $B\rightarrow\rho\pi^{0}$ are varied by $\pm 10\%$.
The uncertainties on the number of $B\overline{B}$ pairs, 
the ratio 
$\tau^{}_{B^+}/\tau^{}_{B^0}$,
and the ratio $f_{+-}/f_{00}$~\cite{Belle:2022hka} are all taken into account.

\section{results}
\label{sec:results}
We determine the branching fractions, {\it CP} asymmetry, and isospin asymmetry directly from the fit.
The results are
\begin{linenomath}
 \begin{align}
  {\cal B}\left(B^{+}\rightarrow\rho^{+}\gamma\right) &= \left(13.1^{+2.0 +1.3}_{-1.9 -1.2}\right)\times 10^{-7}\\
  {\cal B}\left(B^{0}\rightarrow\rho^{0}\gamma\right) &= \left(7.5\pm 1.3^{+1.0}_{-0.8}\right)\times 10^{-7}\\
  A_{\it CP}\left(B^{+}\rightarrow\rho^{+}\gamma\right) &= \left(-8.2\pm 15.2^{+1.6}_{-1.2}\right)\%\\
  A_{\rm I}\left(B\rightarrow\rho\gamma\right) &= \left(10.9^{+11.2 +6.8 +3.8}_{-11.7 -6.2 -3.9}\right)\%,
 \end{align}
\end{linenomath}
where the first uncertainty is statistical, the second is systematic, and the third for $A_{\rm I}$ is the uncertainty from $f_{+-}/f_{00}$~\cite{Belle:2022hka}
combined with the uncertainty from the ratio of $B^{+}$ to $B^{0}$ lifetimes.

In summary, we have measured the branching fractions, direct {\it CP} asymmetry, and isospin asymmetry of $B\rightarrow\rho\gamma$ decays
using a total of 1073~${\mathrm {fb}^{-1}}$ of Belle and Belle~II data.
These results are the most precise to date and supersede the previous Belle results~\cite{Belle:2005grh}.
The results for $A_{\it CP}$ and $A_{\rm I}$ 
are consistent with SM expectations.
\\

This work, based on data collected using the Belle II detector, which was built and commissioned prior to March 2019,
and data collected using the Belle detector, which was operated until June 2010,
was supported by
Higher Education and Science Committee of the Republic of Armenia Grant No.~23LCG-1C011;
Australian Research Council and Research Grants
No.~DP200101792, 
No.~DP210101900, 
No.~DP210102831, 
No.~DE220100462, 
No.~LE210100098, 
and
No.~LE230100085; 
Austrian Federal Ministry of Education, Science and Research,
Austrian Science Fund
No.~P~31361-N36
and
No.~J4625-N,
and
Horizon 2020 ERC Starting Grant No.~947006 ``InterLeptons'';
Natural Sciences and Engineering Research Council of Canada, Compute Canada and CANARIE;
National Key R\&D Program of China under Contract No.~2022YFA1601903,
National Natural Science Foundation of China and Research Grants
No.~11575017,
No.~11761141009,
No.~11705209,
No.~11975076,
No.~12135005,
No.~12150004,
No.~12161141008,
and
No.~12175041,
and Shandong Provincial Natural Science Foundation Project~ZR2022JQ02;
the Czech Science Foundation Grant No.~22-18469S;
European Research Council, Seventh Framework PIEF-GA-2013-622527,
Horizon 2020 ERC-Advanced Grants No.~267104 and No.~884719,
Horizon 2020 ERC-Consolidator Grant No.~819127,
Horizon 2020 Marie Sklodowska-Curie Grant Agreement No.~700525 ``NIOBE''
and
No.~101026516,
and
Horizon 2020 Marie Sklodowska-Curie RISE project JENNIFER2 Grant Agreement No.~822070 (European grants);
L'Institut National de Physique Nucl\'{e}aire et de Physique des Particules (IN2P3) du CNRS
and
L'Agence Nationale de la Recherche (ANR) under grant ANR-21-CE31-0009 (France);
BMBF, DFG, HGF, MPG, and AvH Foundation (Germany);
Department of Atomic Energy under Project Identification No.~RTI 4002,
Department of Science and Technology,
and
UPES SEED funding programs
No.~UPES/R\&D-SEED-INFRA/17052023/01 and
No.~UPES/R\&D-SOE/20062022/06 (India);
Israel Science Foundation Grant No.~2476/17,
U.S.-Israel Binational Science Foundation Grant No.~2016113, and
Israel Ministry of Science Grant No.~3-16543;
Istituto Nazionale di Fisica Nucleare and the Research Grants BELLE2;
Japan Society for the Promotion of Science, Grant-in-Aid for Scientific Research Grants
No.~16H03968,
No.~16H03993,
No.~16H06492,
No.~16K05323,
No.~17H01133,
No.~17H05405,
No.~18K03621,
No.~18H03710,
No.~18H05226,
No.~19H00682, 
No.~20H05850,
No.~20H05858,
No.~22H00144,
No.~22K14056,
No.~22K21347,
No.~23H05433,
No.~26220706,
and
No.~26400255,
and
the Ministry of Education, Culture, Sports, Science, and Technology (MEXT) of Japan;  
National Research Foundation (NRF) of Korea Grants
No.~2016R1\-D1A1B\-02012900,
No.~2018R1\-A2B\-3003643,
No.~2018R1\-A6A1A\-06024970,
No.~2019R1\-I1A3A\-01058933,
No.~2021R1\-A6A1A\-03043957,
No.~2021R1\-F1A\-1060423,
No.~2021R1\-F1A\-1064008,
No.~2022R1\-A2C\-1003993,
and
No.~RS-2022-00197659,
Radiation Science Research Institute,
Foreign Large-Size Research Facility Application Supporting project,
the Global Science Experimental Data Hub Center of the Korea Institute of Science and Technology Information
and
KREONET/GLORIAD;
Universiti Malaya RU grant, Akademi Sains Malaysia, and Ministry of Education Malaysia;
Frontiers of Science Program Contracts
No.~FOINS-296,
No.~CB-221329,
No.~CB-236394,
No.~CB-254409,
and
No.~CB-180023, and SEP-CINVESTAV Research Grant No.~237 (Mexico);
the Polish Ministry of Science and Higher Education and the National Science Center;
the Ministry of Science and Higher Education of the Russian Federation
and
the HSE University Basic Research Program, Moscow;
University of Tabuk Research Grants
No.~S-0256-1438 and No.~S-0280-1439 (Saudi Arabia);
Slovenian Research Agency and Research Grants
No.~J1-9124
and
No.~P1-0135;
Ikerbasque, Basque Foundation for Science, 
the State Agency for Research of the Spanish Ministry of Science and Innovation through Grant No. PID2022-136510NB-C33, 
Agencia Estatal de Investigacion, Spain
Grant No.~RYC2020-029875-I
and
Generalitat Valenciana, Spain
Grant No.~CIDEGENT/2018/020;
the Swiss National Science Foundation; 
National Science and Technology Council,
and
Ministry of Education (Taiwan);
Thailand Center of Excellence in Physics;
TUBITAK ULAKBIM (Turkey);
National Research Foundation of Ukraine, Project No.~2020.02/0257,
and
Ministry of Education and Science of Ukraine;
the U.S. National Science Foundation and Research Grants
No.~PHY-1913789 
and
No.~PHY-2111604, 
and the U.S. Department of Energy and Research Awards
No.~DE-AC06-76RLO1830, 
No.~DE-SC0007983, 
No.~DE-SC0009824, 
No.~DE-SC0009973, 
No.~DE-SC0010007, 
No.~DE-SC0010073, 
No.~DE-SC0010118, 
No.~DE-SC0010504, 
No.~DE-SC0011784, 
No.~DE-SC0012704, 
No.~DE-SC0019230, 
No.~DE-SC0021274, 
No.~DE-SC0021616, 
No.~DE-SC0022350, 
No.~DE-SC0023470; 
and
the Vietnam Academy of Science and Technology (VAST) under Grants
No.~NVCC.05.12/22-23
and
No.~DL0000.02/24-25.

These acknowledgements are not to be interpreted as an endorsement of any statement made
by any of our institutes, funding agencies, governments, or their representatives.

We thank the SuperKEKB team for delivering high-luminosity collisions;
the KEK cryogenics group for the efficient operation of the detector solenoid magnet;
the KEK Computer Research Center for on-site computing support; the NII for SINET6 network support;
and the raw-data centers hosted by BNL, DESY, GridKa, IN2P3, INFN, PNNL/EMSL, and the University of Victoria. 


\begin{thebibliography}{99}
\bibitem{charge-conjugate}
The inclusion of the charge-conjugate modes is implied throughout this paper unless otherwise stated.
\end{thebibliography}
\end{document}